\documentclass[12pt]{iopart}
\usepackage{iopams}
\usepackage[dvipdfm]{graphicx}
\usepackage{bbm}

\newtheorem{lemma}{Lemma}
\newtheorem{theorem}{Theorem}

\begin{document}

\title[Ground states of dipole rings]
{Ground states of classical magnetic dipole rings}
\author{Heinz-J\"urgen Schmidt$^1$
}
\address{$^1$Department of Physics, University of Osnabr\"uck,
 D - 49069 Osnabr\"uck, Germany }

\ead{hschmidt@uos.de}
\vspace{10pt}

\begin{abstract}
We investigate the two well-known ground states of rings of $N$ classical magnetic dipoles that are given by clockwise or anti-clockwise spin orientations tangent to the circle encompassing the dipole ring. In particular, we formulate a rigorous proof of the ground state property of the states in question. The problem can be reduced to the determination of the lowest eigenvalue of a $3N\times 3N$ matrix ${\mathbf J}$. We show that all eigenvalues of ${\mathbf J}$ can be analytically calculated and, at least for  $N=3,\ldots,8$, the lowest one can be directly determined. The main part of the paper is devoted to the completion of the proof for $N\ge 9$ based on various estimates and case distinctions. We also discuss the question
to what extent computer-algebraic results should be allowed to contribute to a mathematical proof.
\end{abstract}

%
\vspace{2pc}

%
%
%
%

\maketitle

\section{Introduction}\label{sec:I}

The ground state(s) of a spin system are important since they determine its behavior and properties at low temperatures. Although at low temperatures quantum
fluctuations are prominent yet the classical ground state(s) may contain valuable information. These classical ground states can be calculated by
numerical or analytical methods for a number of special cases but a general theory does not exist despite a few attempts towards general results
as, e.~g.~, \cite{LT46} or  \cite{SL03}. In this paper we consider a special class of anisotropic systems, namely classical magnetic
dipoles located at the vertices of a regular $N-$polygon interacting via their magnetic fields, in short: dipole rings. There exists an overwhelming numerical evidence that the ground states of dipole rings are given by the two clockwise or anti-clockwise spin orientations tangent to the circle encompassing the dipole ring, see Figure \ref{figGS}. These states, denoted by $\pm{\mathbf t}$,
are consequently assumed to be the ground states in a couple of publications, e.~g.~, \cite{JK95} -- \cite{SSL16}. Nevertheless, a rigorous proof of this fact seems not to exist. The motivation to publish
such a proof is not to dispel remaining doubts about the ground state property of $\pm{\mathbf t}$ but rather to illustrate the problems that may occur
even in analytically solvable cases and to provide methods to cope with these problems.

In section \ref{sec:R} we shortly provide the basic definitions for magnetic dipole rings. For more details the reader is referred to \cite{SSL16} and \cite{SSHL15}.
The proof of the ground state property of the two tangential states $\pm{\mathbf t}$
requires a large number of mostly elementary steps that are, however, intricately intertwined.
In order to help the reader to keep track of the structure of the proof it will be in order to sketch the main ideas
without going into details.

The Hamiltonian of the dipole ring is a bilinear function of the $3N$ spin components $s_{\mu,i}$ and hence can be represented
by a real, symmetric  $3N\times 3N$ matrix ${\mathbf J}$ such that the energy $E$ has the form of an ``expectation value"
$E=\langle{\mathbf s}|{\mathbf J}|{\mathbf s}\rangle$. It turns out that in our case ${\mathbf J}$ can be completely diagonalized
due to the $C_N-$ symmetry of the dipole ring.
In particular, there exists an eigenvector ${\mathbf t}$ of ${\mathbf J}$ that can be identified with
the conjectured tangential ground state of the dipole ring. It ``only" remains to show that the corresponding eigenvalue
$ \frac{\langle{\mathbf t}|{\mathbf J}|{\mathbf t}\rangle}{\langle{\mathbf t}|{\mathbf t}\rangle}$
is the lowest eigenvalue $j_{min}(N)$ of ${\mathbf J}$ for all $N\ge 3$. In this way we have reduced the ground state problem
to a matrix problem in close analogy to the Luttinger-Tisza approach \cite{LT46}.
The matrix in question is even diagonalized in closed form and hence the problem should be tractable.

Actually, the ground state problem could be solved along these lines of thought for any given $N$ provided that it could be treated
by hand or by computer-algebraic software. We have done this in section \ref{sec:A} for $N=3,\ldots,8$.
The challenging problem is rather to prove $j_{min}(N)=\frac{\langle{\mathbf t}|{\mathbf J}|{\mathbf t}\rangle}{\langle{\mathbf t}|{\mathbf t}\rangle}$ for {\it all} $N$. It can be further reduced to the problem whether the
determinant of a $2\times 2$ matrix $K^{(\nu)}$ is strictly positive, where $\nu=1,\ldots,N-1$ is a certain discrete parameter (number of the finite Fourier coefficient or wave number).
However, the matrix entries are not explicitly given numbers but sums of approximately $N$
trigonometric functions and hence the positivity of the determinant is not obvious.

As for many other problems it seems to be a good strategy to look at special cases. As noted above, the special case of small $N$ is well understood. What about the special case of $N\longrightarrow\infty$ ? This leads to the strategy of evaluating the mentioned sums asymptotically, i.~e.~, in
the leading order w.~r.~t.~$N$. The next step, see section \ref{sec:P},
would then be to replace the asymptotic reasoning by strict inequalities and thus to obtain a proof
of the ground state property of $\pm{\mathbf t}$ that is valid for, say, $N\ge N_\ast$. Here we encounter the next complication: Already the asymptotic
reasoning, and the more the formulation of rigorous estimates heavily depends on case distinctions. To explain this we remark that the
above-mentioned determinant can be written as a double sum of the form $\det K^{(\nu)}=\sum_{\lambda,\mu=1}^{N-1}k^{(\nu)}_{\lambda,\mu}$.
An obvious attempt to control the sign of the determinant is to split the terms $k^{(\nu)}_{\lambda,\mu}$ into two parts, say,
$k^{(\nu)}_{\lambda,\mu}={\mathcal P}^{(\nu)}_{\lambda,\mu}+{\mathcal N}^{(\nu)}_{\lambda,\mu}$ :
The first part ${\mathcal P}^{(\nu)}_{\lambda,\mu}$ will be
strictly positive whereas the second part ${\mathcal N}^{(\nu)}_{\lambda,\mu}$  may be positive or negative, depending on the parameters.

One is then looking for a lower bound $B_L$ of the sum of the positive parts and an upper bound $B_U$ of the absolute value of the sum of the possibly negative parts and tries to show $B_L>B_U$ which is sufficient to complete the proof. Unfortunately, the form of the splitting depends on $\lambda$,
more precisely, whether $1\le\lambda\le N/4$ or $N/4\le\lambda\le N/2$ (the other domains of $\lambda$ being reduced to the former ones by means of symmetry arguments). Moreover, the form of the bounds $B_L$ and $B_U$, restricted to partial sums, depends on $\lambda,\mu$ and $\nu$ and we are led to a variety of case distinctions, see table \ref{Tab2}. Especially, the domain of the parameter $\nu$ has to be divided into three parts. One reason
is that, in the case of $1\le\lambda,\mu\le N/4$, the lower bound of the positive terms $B_L$ is of order $O(N^4)$. Without any restriction of $\nu$
we could only show that $B_U=O(N^4)$ which would not be sufficient even for an asymptotic proof. Only with the restriction $\nu\le \varepsilon\sqrt{N}$
we could achieve the result $B_U=O(N^3 (\log N)^2)$ which entails $B_L>B_U$ for $N\ge N_\ast$.
In a similar way we found it necessary to introduce the
further sub-division $ \varepsilon\sqrt{N}<\nu<\delta N$ and $\delta N\le \nu\le \frac{N}{2}$ in order to obtain reasonable bounds. Here the three real
parameters $\varepsilon, \delta$ and $N_\ast$ are first considered as variables and only after all bounds have been established, have to be chosen in an optimal way to obtain an $N_\ast$ as small as possible. Some estimates crucially depend on the assumption $N\ge 9$; hence we introduce the variable $N_0$ that is used in order
to make some equations more transparent but has nevertheless the constant value $N_0=9$. Our final aim is to show $N_\ast=N_0$ but we must not assume
this from the outset. One intermediate attempt gave an $N_\ast=9.04\ldots$ which is slightly above the desirable value of $N_\ast=9$. Hence we had to improve the bounds $B_U$ by using the explicit result $\sum_{\lambda=1}^{N-1}\csc^2\frac{\pi\lambda}{N}=\frac{N^2-1}{3}$ that can be found, e.~g.~, in \cite{H75}, thereby partly abolishing the former case distinction. After this move the choice $\varepsilon=\frac{26}{53},\,\delta=\frac{8}{41}$ was happily consistent with $N_\ast=9$ and I stopped improving estimates. However, the latter result was at first only obtained by numerical means.

Here we face a subtle problem connected with the question to what extent a mathematical proof is allowed to be based on computer-algebraic means.
It is clear that the present proof practically would not be possible without such means. But the crucial question is which means can be tolerated without vitiating the standard of a mathematical proof. I will adopt the position that computer-algebraic means are admissible as long as they could be
replaced by paper-and-pencil operations albeit long and arduous ones. This needs some clarification. Any computer program can, in principle, be simulated by a suitable Turing machine and hence by paper-and-pencil operations. Thus the above definition of ``admissible computer-algebraic means" only makes sense if it is
{\it not} understood to be applicable ``in principle"  but rather ``in practice", even though this introduces some vagueness into the definition.

For example, the simplifications leading from (\ref{A5a}) to the results of
table \ref{Tabev} have been performed using the computer-algebra software MATHEMATICA and not been checked by hand. But this appears to be a harmless use of computer software since it is completely clear what the equivalent paper-and-pencil operations would be and that they could be performed in a reasonable time. On the other hand, the claim that $B_L>B_U$ for $N\ge 9$ is a statement about infinitely many real numbers and cannot be justified by an inspection of a numerically produced graph. The latter cannot be considered as a proof but at most as a strong numerical evidence. Otherwise already the statement $\det K^{(\nu )}>0$ for all $N\ge 3$ and all $1\le \nu \le N-1$ could be ``proven" by inspection of a graph and the whole proof presented in this paper would be pointless.

Given this restriction of the use of computer-algebraic means in a mathematical proof, how should we then complete the present proof? I will explain the chosen strategy for the situation given in Figure \ref{FIGib}. We replaced the condition $B_L>B_U$ by the equivalent one $N^{-4}\,B_L>N^{-4}\,B_U$
and plotted the two functions of $N$ for $0\le N \le 20$. It is evident that there is an intersection at $N=N_B=8.15728\ldots$ and that
$N^{-4}\,B_L>N^{-4}\,B_U$ for $N>N_B$, but this will not suffice for the proof. What we can prove is that $N^{-4}\,B_L$ is an increasing (linear) function of $N$ and that $N^{-4}\,B_U$ is decreasing with the limits $\frac{1}{18} \left(9\, \zeta(3)+2\right)>0$
for $N\longrightarrow\infty$ and $\infty$ for $N\longrightarrow 0$. (By an {\it increasing} function I always mean a {\it strictly monotonically increasing}
one throughout this paper, analogously for {\it decreasing} functions.)
 The graphs of both functions  hence intersect at a unique point with $N=N_B$ and
$N^{-4}\,B_L>N^{-4}\,B_U$ for $N>N_B$. For our purposes we need the stronger result that $N_B<9$ and this will be obtained by means of MATHEMATICA.
This use of computer-algebraic software is now legitimate since it only involves the approximate evaluation of two elementary functions for two arguments, say, $N=5$ and $N=9$. The rationale for allowing computer-algebraic means here is that the approximate evaluation of elementary functions could be done by hand and would yield the same results and only require more time. One might object that the possibility of errors in applying the computer-algebraic software or even in the software itself cannot be excluded but this is beside the point since even an alleged traditional mathematical proof may contain errors.

The remaining part of the proof follows these guidelines.

\section{Rings of interacting magnetic dipoles}
\label{sec:R}
We consider systems of $N$ classical point-like dipoles. The normalized dipole
moments are described by unit vectors ${\mathbf s}_\nu,\;\nu=0,\ldots,N-1$.
Each dipole moment performs a precession about the momentary magnetic field vector
that results as a sum over all magnetic fields produced by the other dipoles.
The $N$ dipoles are fixed at the positions of the vertices of a regular $N-$ polygon
\begin{equation}\label{R1}
{\mathbf r}_\nu =
\left(
\begin{array}{c}
 \cos \frac{2\pi\nu}{N} \\
 \sin \frac{2\pi\nu}{N} \\
 0\\
\end{array}
\right),\quad \nu=0,\ldots,N-1
\;.
\end{equation}
For the sake of simplicity, the length of the vectors ${\mathbf r}_\nu$ is chosen as $1$, but it can be scaled arbitrarily.
The dimensionless energy of the dipole system is
\begin{equation}\label{R4}
H=\sum_{\begin{array}{c}\mu,\nu=0,\ldots N-1\\ \mu\neq \nu\end{array}}
\frac{1}{|{\mathbf r}_\mu-{\mathbf r}_\nu|^3}\;
{\mathbf s}_\nu\cdot
\left( {\mathbf s}_\mu-3\,{\mathbf s}_\mu\cdot {\mathbf e}_{\mu \nu}\;{\mathbf e}_{\mu \nu}
\right)
\;,
\end{equation}
see \cite{G99} (6.35) and \cite{SSHL15}. Here ${\mathbf e}_{\mu \nu}$ denotes the unit vector pointing from the $\nu$-th dipole to the $\mu$-th one:
\begin{equation}\label{defemunu}
   {\mathbf e}_{\mu \nu}\equiv\frac{{\mathbf r}_\mu-{\mathbf r}_\nu}{|{\mathbf r}_\mu-{\mathbf r}_\nu|}\;.
\end{equation}

\begin{figure}
\begin{center}
\includegraphics[clip=on,width=120mm,angle=0]{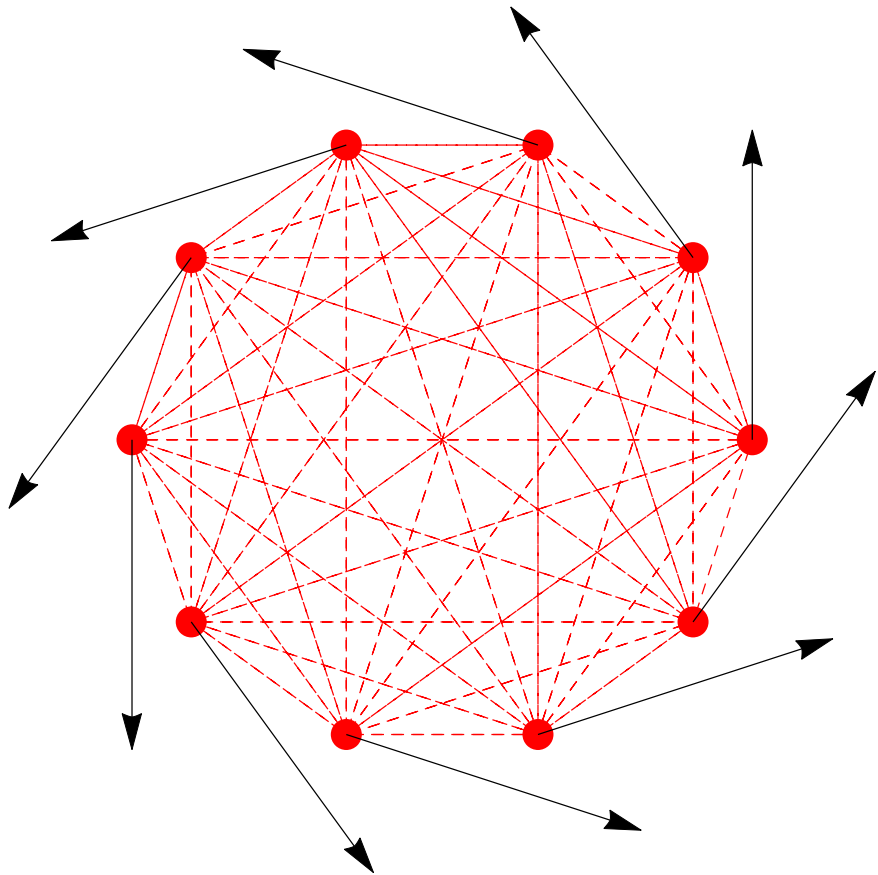}
\end{center}
\caption{Illustration of one of the two ground states $\pm {\mathbf t}$ of the $N=10$ dipole ring.
\label{figGS}
}
\end{figure}

By definition, the {\it ground states} of the system are spin configurations that minimize
the energy (\ref{R4}). Numerical studies suggest that there are exactly two ground states, namely
\begin{equation}\label{R5}
{\mathbf t}_\nu=\left(
\begin{array}{c}
 -\sin \frac{2\pi\nu}{N} \\
 \cos \frac{2\pi\nu}{N} \\
 0
\end{array}
\right),\quad \nu=0,\ldots,N-1
\;,
\end{equation}
and $-{\mathbf t}_\nu,\;\nu=0,\ldots,N-1$, see figure \ref{figGS} for an illustration.
The present paper is devoted to the proof of this fact.

\section{Ground states of the dipole ring}\label{sec:A}

Obviously, the Hamiltonian (\ref{R4}) is bilinear in the components $s_{\mu i}$ of the moment vectors
${\mathbf s}_\mu$
and hence can be written in the form
\begin{eqnarray}\label{A1}
H&=&\sum_{\mu=0}^{N-1}\sum_{\nu=0}^{N-1}\sum_{i,j=1}^3 J_{\mu\nu i j} s_{\mu i}s_{\nu j}\\
&\equiv& \sum_{\alpha,\beta} {\mathbf J}_{\alpha\beta} s_\alpha s_\beta
\;,
\end{eqnarray}
where we have introduced multi-indices $\alpha=(\mu,i)\;,\beta=(\nu,j)$ that run through a finite set of size $3N$.
Let $j_{\mbox{\scriptsize min}}$ be the lowest eigenvalue of the symmetric matrix ${\mathbf J}$. Then,
by the Rayleigh-Ritz variation principle,
$H\ge \sum_{\alpha} j_{\mbox{\scriptsize min}} s_\alpha^2=N j_{\mbox{\scriptsize min}}$,
but the minimal energy $E_0$ need not be equal to $N j_{\mbox{\scriptsize min}}$ in general. We will prove that
there exists a certain eigenvalue $j_\alpha$ of ${\mathbf J}$ such that the corresponding eigenvector
can be identified with the state ${\mathbf t}$, see (\ref{R5}). Hence ${\mathbf t}$ is a ground state
if $j_\alpha=j_{min}$ since in this case the lower bound $N\,j_{min}$ of the energy is assumed by the spin configuration ${\mathbf t}$.\\
To detail the above remarks it is convenient to introduce new cartesian coordinates
$(\xi,\eta,\zeta)\equiv(\xi_0,\ldots,\xi_{N-1},\eta_0,\ldots,\eta_{N-1},\zeta_0,\ldots,\zeta_{N-1})$
for the moment vectors ${\mathbf s}_\mu$ that are better adapted to the $C_N$-symmetry of the problem:
\begin{equation}\label{A2}
\xi_\mu={\mathbf r}_\mu\cdot{\mathbf s}_\mu,\;\eta_\mu={{\mathbf t}}_\mu\cdot{\mathbf s}_\mu,\;,
\zeta_\mu={\mathbf e}\cdot{\mathbf s}_\mu,\quad \mu=0,\ldots,N-1\;,
\end{equation}
where ${\mathbf e}\equiv\left(0,0,1\right)^\top$.
 In the following we will express the energy $H$ in terms of the new coordinates (\ref{A2}),
where the transformed matrix will again be denoted by  ${\mathbf J}$ without danger of confusion.
To this end we consider the part of the energy that is linear in ${\mathbf s}_0$:
\begin{equation}\label{H0}
 H_0\equiv \sum_{\mu\neq 0} \frac{1}{|{\mathbf r}_0-{\mathbf r}_\mu|^3}\;\left(
{\mathbf s}_0\cdot{\mathbf s}_\mu-3\,{\mathbf s}_0\cdot {\mathbf e}_{0\mu}\;{\mathbf s}_\mu\cdot {\mathbf e}_{0\mu}
\right)
\;.
\end{equation}
For the intermediate steps of the calculation we set $c\equiv \cos\frac{2\pi\mu}{N}$ and $s\equiv \sin\frac{2\pi\mu}{N}$.
After elementary transformations we obtain
\begin{eqnarray}\label{A3a}
  {\mathbf r}_0-{\mathbf r}_\mu &=& \left(\begin{array}{c}1-c \\-s \\ 0\end{array}\right)\;, \\
  \label{A3b}
  \left|{\mathbf r}_0-{\mathbf r}_\mu\right|^2 &=& (1-c)^2+s^2=2(1-c) =4\sin^2\frac{\pi\mu}{N}\;, \\
   \label{A3c}
  \frac{1}{ \left|{\mathbf r}_0-{\mathbf r}_\mu\right|^3}&=& \frac{1}{8}\csc^3 \frac{\pi\mu}{N}\;,\\
   \label{A3d}
   {\mathbf s}_0\cdot{\mathbf s}_\mu&=& \xi_0\xi_\mu\,c-\xi_0\eta_\mu\,s+\eta_0\xi_\mu\, s+\eta_0\eta_\mu c +\zeta_0\zeta_\mu\;,\\
   \label{A3e}
   -3\,{\mathbf s}_0\cdot {\mathbf e}_{0\mu}\;{\mathbf s}_\mu\cdot {\mathbf e}_{0\mu}
   &=&
   \frac{3}{2}\left[
   \xi_0\xi_\mu\,(1-c)+\xi_0\eta_\mu\,s-\eta_0\xi_\mu\, s-\eta_0\eta_\mu (1+ c)    \right]\;,\\
   \nonumber
    {\mathbf s}_0\cdot{\mathbf s}_\mu -3\,{\mathbf s}_0\cdot {\mathbf e}_{0\mu}\;{\mathbf s}_\mu\cdot {\mathbf e}_{0\mu}
   &=&
   \frac{1}{2}\left[
   \xi_0\xi_\mu\,(3-c)+\xi_0\eta_\mu\,s-\eta_0\xi_\mu\, s-\eta_0\eta_\mu (3+ c)+\zeta_0\zeta_\mu \right]\;.
   \\&&\label{A3f}
\end{eqnarray}
From these equations one can read off the first, the $N+1$-th and the $2N+1$-th row (and the analogous columns) of the matrix ${\mathbf J}$.
The other rows can be obtained by cyclic permutations of $(0,1,2,\ldots, N-1)$. More precisely,
the matrix ${\mathbf J}$ assumes the form
\begin{equation}\label{A3}
{\mathbf J}=
\left(
\begin{array}{ccc}
 A&C&0 \\
 -C&B&0 \\
 0&0&D \\
\end{array}
\right)
\;,
\end{equation}
where $A,B,C,D$ denote $N\times N$ sub-matrices that are so-called circulants, see \cite{A01}.
A {\it circulant} is an $N\times N$-matrix that commutes with the cyclic permutation matrix of $(0,1,2,\ldots, N-1)$.
As an example we display the sub-matrix $A$ for $N=4$:
\begin{equation}\label{Ac}
A=\left(
\begin{array}{cccc}
0 &\frac{3}{8 \sqrt{2}}&\frac{1}{8}&\frac{3}{8 \sqrt{2}} \\
\frac{3}{8 \sqrt{2}}&0 &\frac{3}{8 \sqrt{2}}&\frac{1}{8} \\
\frac{1}{8}&\frac{3}{8 \sqrt{2}}&0 &\frac{3}{8 \sqrt{2}} \\
\frac{3}{8 \sqrt{2}}&\frac{1}{8}&\frac{3}{8 \sqrt{2}}&0  \\
\end{array}
\right)
\;.
\end{equation}
One notes that $A$ has constant secondary diagonals even if these are periodically extended.
The eigenvectors ${\mathbf b}^{(\mu)}$ of a circulant form the Fourier basis, i.~e.~, are of the form
\begin{equation}\label{L3a}
{\mathbf b}^{(\mu)}_\nu =\frac{1}{\sqrt{N}}\exp\left(
\frac{2\pi\,i\,\mu\,\nu}{N}
\right),
\quad\mu,\nu=0,\ldots,N-1
\;,
\end{equation}
and the eigenvalues are the Fourier transform (times $\sqrt{N}$) of the circulant's first row, see \cite{A01}.\\
$A,B$ and $D$ are symmetric, whereas $C$ is anti-symmetric. The matrices $A,B,C,D$
pairwise commute since they have the Fourier basis (\ref{L3a}) as a common system of eigenvectors.
Since they are circulants it suffices to give the entries of the first row of the respective matrices.
These values can be read off from (\ref{A3c}) and (\ref{A3f}):
\begin{eqnarray}\label{A4a}
A_{0,\mu}&=&
\left\{
\begin{array}{l@{\;:\;}l}
 0& \mu=0\,,\\
\frac{1}{32} \left(3-\cos \left(\frac{2 \pi  \mu }{N}\right)\right) \csc ^3\left(\frac{\pi  \mu
   }{N}\right)
 & \mu=1,\ldots,N-1,
\end{array}
\right.\\
\label{A4b}
B_{0,\mu}&=&
\left\{
\begin{array}{l@{\;:\;}l}
 0& \mu=0\,,\\
-\frac{1}{32} \left(3+\cos \left(\frac{2 \pi  \mu }{N}\right)\right) \csc ^3\left(\frac{\pi  \mu
   }{N}\right)
 & \mu=1,\ldots,N-1,
\end{array}
\right.\\
\label{A4c}
C_{0,\mu}&=&
\left\{
\begin{array}{l@{\;:\;}l}
 0& \mu=0\,,\\
\frac{1}{32} \sin \left(\frac{2 \pi  \mu }{N}\right) \csc ^3\left(\frac{\pi  \mu }{N}\right)
 & \mu=1,\ldots,N-1,
\end{array}
\right.
\\
\label{A4d}
D_{0,\mu}&=&
\left\{
\begin{array}{l@{\;:\;}l}
 0& \mu=0\,,\\
\frac{1}{16} \csc ^3\left(\frac{\pi  \mu }{N}\right)
 & \mu=1,\ldots,N-1,
\end{array}
\right.
\end{eqnarray}

We note that, except a vanishing diagonal, $A$ and $D$  have only positive entries and $B$ has only negative ones.
The row sum of $C$ vanishes since $C$ is an anti-symmetric circulant. Moreover, the eigenvalues of $C$ are purely imaginary
since it is also anti-Hermitian. (From this it follows again that the row sum, which is a real eigenvalue of $C$, must vanish.)
Let
\begin{equation}\label{Jmat}
  J^{(\mu)}=\left(
\begin{array}{ccc}
 a^{(\mu)}& {\sf i}\,c^{(\mu)}&0 \\
 -{\sf i}\,c^{(\mu)}&b^{(\mu)}&0 \\
 0&0&d^{(\mu)} \\
\end{array}
\right)
\end{equation}
be the  $3\times 3$ matrix where $a^{(\mu)},b^{(\mu)}, {\sf i}\,c^{(\mu)},d^{(\mu)} $ are the eigenvalues of the corresponding sub-matrices
$A,B,C,D$
of ${\mathbf J}$ and $\mu=0,\ldots,N-1$.
Further let $j^{(\mu)}_i,\;i=1,2,3$ be the eigenvalues of $J^{(\mu)}$
with eigenvectors $u^{(\mu)}_i$.
Then the general eigenvector of $\mathbf J$ has the form
$(u^{(\mu)}_{i,1}{\mathbf b}^{(\mu)},u^{(\mu)}_{i,2}{\mathbf b}^{(\mu)},u^{(\mu)}_{i,3}{\mathbf b}^{(\mu)})^\top$
corresponding to the eigenvalue  $j^{(\mu)}_i$.

 In this way we have, in principle, diagonalized the matrix ${\mathbf J}$. In particular, its eigenvalues corresponding to $\mu=0$
 can be determined explicitely. The Fourier basis vector ${\mathbf b}^{(0)}$ is the vector with constant entries $\frac{1}{\sqrt{N}}$.
 The eigenvalues $a^{(0)}, b^{(0)}, {\sf i}\,c^{(0)}, d^{(0)}$ considered above are the constant row sums of $A,B,C,D$, where $c^{(0)}=0$, since $C$ has
 vanishing row sums. It follows that $J^{(0)}=\mbox{diag }(a^{(0)}, b^{(0)},  d^{(0)})$. Obviously, $b^{(0)}$ is the lowest eigenvalue of $J^{(0)}$
 since $b^{(0)}<0$ but $a^{(0)}>0$ and $d^{(0)}>0$. The corresponding eigenvector of ${\mathbf J}$ is
 $({\mathbf 0},{\mathbf b}^{(0)},{\mathbf 0})^\top$.
 It is, up to normalization, identical with the conjectured ground state ${\mathbf t}$ according to (\ref{R5}). To prove that
 ${\mathbf t}$ is actually a ground state it would suffice to show that $b^{(0)}$ is the lowest eigenvalue of ${\mathbf J}$
 since then the equality sign in $E_0\ge N j_{min}$ would be assumed.\\

\begin{figure}
\begin{center}
\includegraphics[clip=on,width=150mm,angle=0]{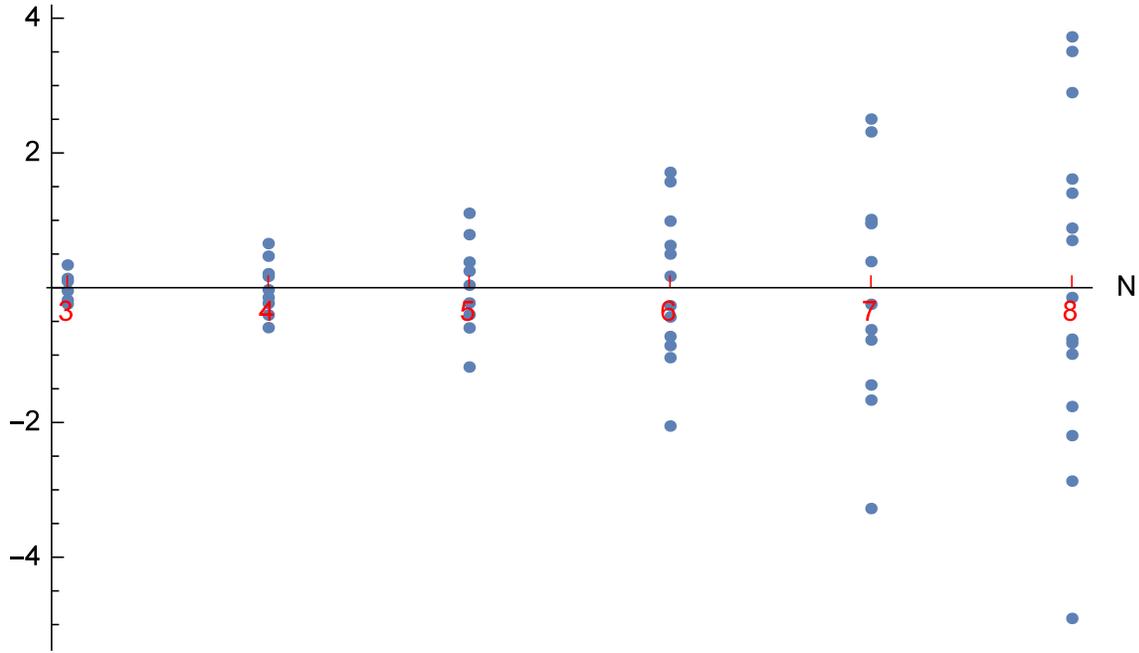}
\end{center}
\caption{Eigenvalues of the matrix ${\mathbf J}$, see (\ref{A3}) and (\ref{A4a})--(\ref{A4d}), for $N=3,\ldots,8$. The lowest eigenvalue $j_{min}=b^{(0)}$
corresponding to the ground state energy $E_0=N\,j_{min}$
can be analytically calculated, see table \ref{Tabev}.
\label{FIGev}
}
\end{figure}

Recall that
\begin{eqnarray}\label{A5a}
b^{(0)}  &=&\sum_{\mu=0}^{N-1}B_{0,\mu}\stackrel{(\ref{A4b})}{=}-\frac{1}{32}
\sum_{\mu=1}^{N-1}
\left(3+\cos \frac{2 \pi  \mu }{N}\right) \csc ^3\frac{\pi  \mu   }{N}\\
\label{A5b}
&=&-\frac{1}{16}
\sum_{\mu=1}^{N-1}
\left(1+\cos^2 \frac{ \pi  \mu }{N}\right) \csc ^3\frac{\pi  \mu   }{N}
\;.
\end{eqnarray}
For small $N$ there exists an even simpler analytical form of $b^{(0)}$, see table \ref{Tabev}, except for
$N=7$, where we could only simplify (\ref{A5b}) to
\begin{equation}\label{b7}
 b^{(0)}=\frac{1}{8} \left(-2 \csc ^3\frac{\pi }{7}+\csc \frac{\pi }{7}-2 \sec ^3\frac{\pi }{14}-2
   \sec ^3\frac{3 \pi }{14}+\sec \frac{\pi}{14}+\sec \frac{3 \pi }{14}\right)
   \;.
\end{equation}
Obviously, not only $b^{(0)}$ but all eigenvalues of ${\mathbf J}$ can be calculated in closed form for, say, $N=3,\ldots,8$.
Hence the claim that $j_{min}=b^{(0)}$ can be confirmed for these cases by numerical evaluation of given expressions involving only
elementary functions.

\begin{table}
\caption{\label{Tabev}Table of the analytical form of $j_{min}=b^{(0)}$ for $N=3,\ldots,8$.
\\}
\begin{center}\begin{tabular}{|c|c|c|}\hline\hline
N &$j_{min}$ & numerical value\\ \hline\hline
3& $-\frac{5}{12 \sqrt{3}}$ &$-0.240563$\\ \hline
4& $\frac{1}{16} \left(-1-6 \sqrt{2}\right)$ &$-0.59283$\\ \hline
5& $-\frac{1}{20} \sqrt{425+58 \sqrt{5}}$ &$-1.17759$\\ \hline
6& $-\frac{29}{16}-\frac{5}{12 \sqrt{3}}$ &$-2.05306$\\ \hline
7& (\ref{b7}) &$-3.27741$\\ \hline
8& $\frac{1}{16} \left(-1-6 \sqrt{2}-4 \sqrt{274+17 \sqrt{2}}\right)$ &$-4.9088$\\
\hline\hline
\end{tabular}\end{center}
\end{table}


\section{Confirmation of the ground states for $N\ge 9$}
\label{sec:P}


In this section we will prove the ground state property of ${\mathbf t}$ for sufficiently large $N$,
more precisely, for
\begin{equation}\label{ngn0}
  N\ge N_0\equiv 9
  \;.
\end{equation} This result is
the more plausible since in the limit $N\longrightarrow\infty$ the dipole ring approaches the infinite chain that has, up to a sign, a unique ground
state where all spins are aligned parallel or anti-parallel w.~r.~t.~the chain direction \cite{SSL16}. This ground state minimizes the energy of
every single pair interaction and is hence unfrustrated.\\
First we argue that
\begin{equation}\label{Perron1}
b^{(0)}<d^{(\nu)} \mbox{  for all } \nu=0,\ldots,N-1
\;.
\end{equation}
Recall that $d^{(\nu)}$ can be written as
\begin{eqnarray}\label{A5c}
d^{(\nu)}&=&\sum_{\mu=1}^{N-1}\frac{1}{16} \csc ^3\left(\frac{\pi  \mu }{N}\right) \exp\left(\frac{2\pi{\sf i}\nu\mu}{N} \right)\\
\label{A5d}
&=&\sum_{\mu=1}^{N-1}\frac{1}{16} \csc ^3\left(\frac{\pi  \mu }{N}\right) \cos\left(\frac{2\pi\nu\mu}{N} \right)
\;,
\end{eqnarray}
and hence
\begin{equation}\label{A6}
|d^{(\nu)}|\le \sum_{\mu=1}^{N-1}\frac{1}{16} \csc ^3\left(\frac{\pi  \mu }{N}\right)=d^{(0)}
\;,
\end{equation}
where the $=$ sign only applies for $\nu=0$.
Alternatively, we could have invoked the theorem of Perron (1907) in the form \cite{N76} in order to show (\ref{A6}).
Now we use the fact that $D_{0,\nu}<|B_{0,\nu}|$ for all $\nu=0,\ldots,N-1$ and hence $d^{(0)}<|b^{(0)}|$. Together with (\ref{A6})
this implies $|d^{(\nu)}|<|b^{(0)}|$ and further $b^{(0)}<d^{(\nu)}$ since $b^{(0)}<0$. \\
Similarly one also proves
\begin{equation}\label{Perron2}
b^{(0)}<b^{(\nu)} \mbox{  for all } \nu=0,\ldots,N-1
\;.
\end{equation}
Now we can restrict ourselves to the $2\times 2$ submatrix
$\left(
\begin{array}{cc}
 a^{(\nu)}& {\sf i}\,c^{(\nu)}\\
 -{\sf i}\,c^{(\nu)}&b^{(\nu)} \\
\end{array}
\right)$
of $J^{(\nu)}$.
We subtract $b^{(0)}$ in the diagonal and obtain the matrix
\begin{equation}\label{K}
K^{(\nu)}\equiv
\left(
\begin{array}{cc}
 a^{(\nu)}-b^{(0)}& {\sf i}\,c^{(\nu)}\\
 -{\sf i}\,c^{(\nu)}&b^{(\nu)}-b^{(0)} \\
\end{array}
\right)
\;.
\end{equation}
The ground state property now follows if  $K^{(\nu)}$ is positive-definite,
i.~e.~if both eigenvalues of $K^{(\nu)}$ are strictly positive for all $\nu=0,\ldots,N-1$.
By virtue of Sylvester's criterion (positivity of all principal minors) and the positivity of $K^{(\nu)}_{22}$,
see (\ref{Perron2}), it remains to show that
\begin{equation}\label{det}
 \det K^{(\nu)}=(a^{(\nu)}-b^{(0)})(b^{(\nu)}-b^{(0)})-(c^{(\nu)})^2>0
 \;.
\end{equation}
After some elementary transformations we write $ \det K^{(\nu)}$ as a double sum of the form
\begin{eqnarray}
 \nonumber 
  \det K^{(\nu)} &=& \sum_{\lambda,\mu=1}^{N-1}k^{(\nu)}_{\lambda\mu}\\
  \nonumber
   &\equiv& \sum_{\lambda,\mu=1}^{N-1}
   \csc ^3\frac{2 \pi  \lambda }{N} \csc ^3  \frac{2 \pi  \mu }{N}
   \left[\left( 3+\cos\frac{2\pi\mu}{N}\right)\right.
   \\
  \nonumber
  && \left(1-\cos \frac{2 \pi  \mu  \nu   }{N}\right)
   \left(\cos\frac{2 \pi  \lambda   }{N}
    \left(1-\cos \frac{2 \pi  \lambda  \nu   }{N}\right)+3 \left(\cos \frac{2 \pi  \lambda  \nu}{N}  +1\right)\right)\\
  \label{Adouble}
   &&\left.-\sin \frac{2 \pi  \lambda   }{N}\sin \frac{2 \pi  \mu }{N}\sin \frac{2 \pi  \lambda \nu  }{N}
    \sin  \frac{2 \pi  \mu  \nu }{N}\right]
    \;,
\end{eqnarray}
ignoring the irrelevant global factor $\left(\frac{1}{32}\right)^2$.
It turns out that some terms in the double sum (\ref{Adouble}) are positive and some terms are negative. We have to show that
the positive terms dominate the sum. To this end we will find some lower bound $B_L$ of the sum over all positive terms and
some upper bound $B_U$ of the absolute value of the sum of all negative terms and will show $B_L-B_U>0$ for sufficiently large $N$,
i.~e.~for $N\ge N_\ast$.\\

The terms $k^{(\nu)}_{\lambda,\mu}$ of the double sum (\ref{Adouble}) possess the reflection symmetry
$k^{(\nu)}_{N-\lambda,\mu}=k^{(\nu)}_{\lambda,\mu}=k^{(\nu)}_{\lambda,N-\mu}$. We will utilize this and restrict the
summation to the domain $1\le \lambda,\mu\le N/2$ which yields $1/4$ of the total sum. If $N$ is even we accordingly would have to split
the terms with $\lambda=N/2$ or $\mu=N/2$ into two equal parts belonging to the different partial sums. The total factor of $4$ will be ignored.
Note that for the cases
$BNs$ and $CNs$ considered below it is more convenient to sum over the whole domain and hence a factor $1/4$ is introduced for compensation.\\

The various estimates depend on the values of the parameter $\nu$ (wave number) and of the summation indices $\lambda$ and $\mu$.
This entails a considerable number of case distinctions that are displayed in Table \ref{Tab2}. Due to the symmetry
\begin{equation}\label{nusymm}
k^{(N-\nu)}_{\lambda\mu}=k^{(\nu)}_{\lambda\mu}
\end{equation}
we may restrict ourselves to $1\le\nu\le\frac{N}{2}$.
The constants $\varepsilon$  and $\delta$ occurring in Table \ref{Tab2} are chosen as
\begin{eqnarray}\label{epsilon}
\varepsilon&=& \frac{26}{53}= 0.490566\ldots,\\
\label{delta}
\delta&=&\frac{8}{41}=0.195122\ldots
   \;.
\end{eqnarray}

\begin{table}
\caption{{\label{Tab2}}Table of case distinctions. The letters $P,\,N,\,N_1,\, N_2$ refer to the splitting of
$k^{(\nu)}_{\lambda\mu}$ according to (\ref{split1a}) -- (\ref{split2d}).
The values of $\varepsilon$  and $\delta$ are given in (\ref{epsilon}) and (\ref{delta}).\\}
\begin{center}\begin{tabular}{|c|c|c|}\hline\hline
A & B & C\\ \hline
$1\le \nu\le \nu_A\equiv\varepsilon \sqrt{N}$ & $\nu_A< \nu < \nu_B\equiv\delta N$ & $\nu_B\le \nu \le \frac{N}{2}$\\
\hline
\hline
1 & 2\\ \hline
$1\le \lambda \le \frac{N}{4}$ &$\frac{N}{4}\le \lambda \le \frac{N}{2}$\\
 \hline
$\alpha$ & $\beta$&\\
\hline
$1\le \mu \le \frac{N}{4}$ \mbox{ or } & $\frac{N}{4}\le \mu \le \frac{N}{2}$&\\
\hline\hline
$P$ & $N,\,N_1,\;N_2$&\\ \hline
${\mathcal P}^{(\nu)}_{\lambda\mu}$ & ${\mathcal N}^{(\nu)}_{i,\lambda\mu}$&\\
\hline\hline
\end{tabular}\end{center}
\end{table}

The term $k^{(\nu)}_{\lambda\mu}$ in (\ref{Adouble})
can be split into two parts such that the first one is always positive and only the second one may be negative.
This splitting depends on the sign of $\cos\frac{2 \pi  \lambda   }{N}$, hence on $\lambda$. More precisely, we define\\

Case $1\; (1\le \lambda \le \frac{N}{4} \mbox{ or }\frac{N}{4}\le \lambda \le \frac{3 N}{4})$:\\
\begin{eqnarray}\label{split1a}
k^{(\nu)}_{\lambda\mu}&=& \left(
{\mathcal P}^{(\nu)}_{\lambda\mu} + {\mathcal N}^{(\nu)}_{\lambda\mu}
 \right) \csc^3 \frac{\pi\lambda}{N}\, \csc^3 \frac{\pi\mu}{N}\;,\\
 \nonumber
{\mathcal P}^{(\nu)}_{\lambda\mu} &=&
\left(3+\cos\frac{2\pi\mu}{N} \right)
\left[\cos\frac{2\pi\lambda}{N} \left(1-\cos\frac{2\pi\lambda\nu}{N} \right) +
3  \left(1+\cos\frac{2\pi\lambda\nu}{N} \right)
\right]\\
\label{split1b}
&& \left(1-\cos\frac{2\pi\mu\nu}{N} \right)\;,\\ \label{split1c}
{\mathcal N}^{(\nu)}_{\lambda\mu} &=& - \sin\frac{2\pi\lambda}{N}\, \sin\frac{2\pi\mu}{N}\, \sin\frac{2\pi\lambda\nu}{N}\, \sin\frac{2\pi\mu\nu}{N}
\;.
\end{eqnarray}

Case $2\; (\frac{N}{4}\le \lambda \le \frac{3N}{4} )$:\\
\begin{eqnarray}\label{split2a}
k^{(\nu)}_{\lambda\mu}&=& \left(
{\mathcal P}^{(\nu)}_{\lambda\mu} + {\mathcal N}^{(\nu)}_{1,\lambda\mu} + {\mathcal N}^{(\nu)}_{2,\lambda\mu}
 \right) \csc^3 \frac{\pi\lambda}{N}\, \csc^3 \frac{\pi\mu}{N}\;,\\
\label{split2b}
{\mathcal P}^{(\nu)}_{\lambda\mu} &=&
3\,\left(3+\cos\frac{2\pi\mu}{N} \right)
\left(1+\cos\frac{2\pi\lambda\nu}{N} \right)
\left(1-\cos\frac{2\pi\mu\nu}{N} \right),
\\ \label{split2c}
{\mathcal N}^{(\nu)}_{1,\lambda\mu} &=&
\left(3+\cos\frac{2\pi\mu}{N} \right)
\cos\frac{2\pi\lambda}{N}
\left(1-\cos\frac{2\pi\lambda\nu}{N} \right)
\left(1-\cos\frac{2\pi\mu\nu}{N} \right),
\\ \label{split2d}
{\mathcal N}^{(\nu)}_{2,\lambda\mu} &=& - \sin\frac{2\pi\lambda}{N}\, \sin\frac{2\pi\mu}{N}\, \sin\frac{2\pi\lambda\nu}{N}\, \sin\frac{2\pi\mu\nu}{N}
\;.
\end{eqnarray}

As indicated by the letters ${\mathcal P}$ (positive) and ${\mathcal N}$ (possibly negative) it is easily shown that
${\mathcal P}^{(\nu)}_{\lambda\mu}\ge 0$ whereas the sign of ${\mathcal N}^{(\nu)}_{\lambda\mu}$ or ${\mathcal N}^{(\nu)}_{i,\lambda\mu}$
depends on $\lambda,\mu,\nu$. Note that $\csc \frac{\pi\lambda}{N}$ and $\csc \frac{\pi\mu}{N}$ are always positive since
$0<\lambda,\mu <N$.\\

In the following we calculate the various estimates depending on the case distinctions according to Table \ref{Tab2}.
The notation will be self-explaining; e.~g.~, case A1$\alpha$P means that we investigate the contribution from the positive terms
in the double sum (\ref{Adouble}) corresponding to the summation over $\lambda=1,\ldots,\frac{N}{4}$ and $\mu=1,\ldots,\frac{N}{4}$
where $1\le\nu\le \varepsilon\sqrt{N}$ is assumed.\\

A few words about the summation limits are in order. If $N$ is odd then, e.~g.~, the notation $\sum_{\lambda=N/4}^{N/2}\ldots$
means that the sum has to be performed over all integers in the interval $(N/4,N/2)$. If $N$ is even we have already mentioned the
convention to split the term with $\lambda=N/2$ into two equal parts. If, moreover, $4$ divides $N$ the term with $\lambda=N/4$
has to be assigned not to the sum $\sum_{\lambda=N/4}^{N/2}\ldots$ but to $\sum_{\lambda=1}^{N/4}\ldots$
such that the number of terms in each partial sum never exceeds $N/4$.
This convention simplifies the formulation of estimates like (\ref{A1N5b}). \\

\begin{figure}
\begin{center}
\includegraphics[clip=on,width=150mm,angle=0]{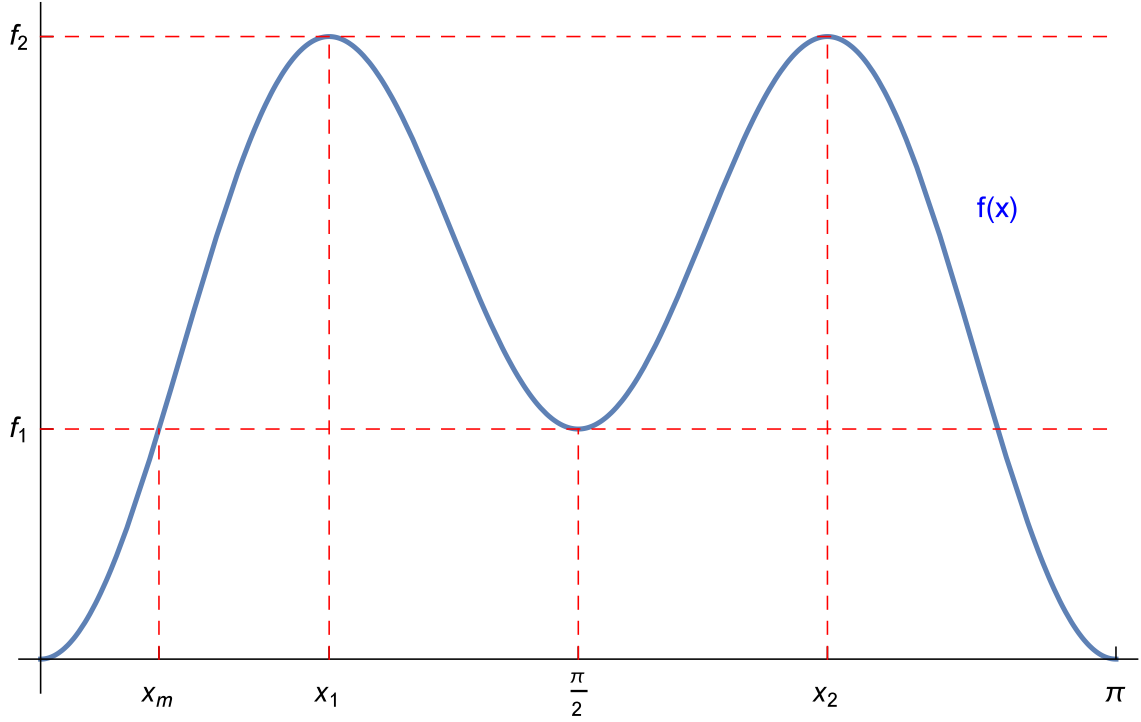}
\end{center}
\caption{Typical form of the function $f(x)$, see (\ref{A1P2}). We chose $N=5$ in order to have a marked local minimum at $x=\pi/2$.
\label{FIGfx}
}
\end{figure}

\fbox{Case A1$\alpha$P}\\

We consider the corresponding part of $\sum_{\lambda\mu}k^{(\nu)}_{\lambda\mu}$ denoted by $K_{A1\alpha P}$. Since all terms of the sum
are positive we obtain a lower bound by restricting the sum to the two terms with $\lambda=\mu=1$ and $\lambda=1,\,\mu=2$:
\begin{equation}\label{A1P1}
 K_{A1\alpha P}\equiv \sum_{\lambda=1}^{N/4}\sum_{\mu=1}^{N/4}
 {\mathcal P}^{(\nu)}_{\lambda\mu} \csc^3 \frac{\pi\lambda}{N}\, \csc^3 \frac{\pi\mu}{N}
 >  {\mathcal P}^{(\nu)}_{11} \csc^6 \frac{\pi}{N}+ {\mathcal P}^{(\nu)}_{12} \csc^3 \frac{\pi}{N}\csc^3 \frac{2\pi}{N}
 \;.
\end{equation}
Let us begin with the first term at the r.~h.~s.~of (\ref{A1P1}) corresponding to $\lambda=\mu=1$.\\
We have to find a lower bound of $ {\mathcal P}^{(\nu)}_{11}$.
With the abbreviations $c\equiv \cos\frac{2\pi}{N}$ and $x\equiv\frac{\pi\,\nu}{N}$ we obtain from (\ref{split1b}), after some simplifications:
\begin{equation}\label{A1P2}
  {\mathcal P}^{(\nu)}_{11}= 4 (3+c) \left(3+(c-3)\sin^2 x\right)\sin^2 x\equiv f(x)
  \;.
\end{equation}
The real function $f(x)$ considered for arguments
$x\in[0,\pi]$ satisfies $f(\pi/2-x)=f(\pi/2+x)$ and has two maxima at $x=x_{1,2}$ and a local minimum at $x=\pi/2$ with height
$f_1=4(3+c)c$, see
Figure \ref{FIGfx}. This follows from
\begin{equation}\label{abl}
\frac{d\,f(x)}{dx}=8 (c+3) \sin (x) \cos (x) \left(2 (c-3) \sin ^2(x)+3\right)=24 (c+3) x+O\left(x^3\right)
\;.
\end{equation}
Hence $\frac{d\,f(x)}{dx}>0$ for sufficiently small values of $x>0$. The derivative (\ref{abl}) vanishes for $x=0,\pi/2,\pi$ and for
\begin{equation}\label{max12}
 \sin^2 x=\frac{3}{2(3-c)}
 \;.
 \end{equation}
This equation has two solutions $x_{1,2}$ such that $0<x_1<\pi/2 < x_2 < \pi$ which yields two maxima of height
\begin{equation}\label{f2}
f_2=f(x_{1,2})=\frac{54}{3-c}-9
 \;.
 \end{equation}
The form of the graph of $f(x)$ given in Figure \ref{FIGfx} is typical since always $f_2>f_1$ due to
\begin{equation}\label{minmax}
f_2-f_1 =\frac{(3-2 c)^2 (c+3)}{3-c}>0
\;.
\end{equation}
Hence $f$ is not  increasing for the whole interval $x\in[0,\pi/2]$ but only for $x\in[0,x_1]$.
Define $x_0\equiv\frac{\pi}{N},\;x_A\equiv\frac{\pi\varepsilon}{\sqrt{N}}$ and $x_B\equiv\delta \pi$ such that $1\le\nu\le \nu_A$ is equivalent to $x_0\le x\le x_A$
and $x_A\le\nu\le \nu_B$ to $x_A\le x\le x_B$.
We note that $x_0<x_A$ for $N>\varepsilon^{-2}=4.15533\ldots$ and $x_A<x_B$ for $N>\frac{\varepsilon^2}{\delta^2}=6.32095\ldots$.
Anticipating the analogous problem of finding a lower bound of ${\mathcal P}^{(\nu)}_{11}$ in the cases B and C we prove
a stronger statement than needed for the present case:
\begin{lemma}\label{Lemmamono}
(i) $f(x_B)<f(\pi/2)$.\\
(ii) If $x_0\le x \le x_A$ then $f(x_0)\le f(x)$.\\
(iii) If $x_A\le x \le x_B$ then $f(x_A)\le f(x)$.\\
(iv) If $x_B\le x \le \frac{\pi}{2}$ then $f(x_B)\le f(x)$.
\end{lemma}
{\bf Proof}:
(i) The real function $N\mapsto \frac{c}{3-c},\;c\equiv\cos\frac{2\pi}{N}$ is  increasing for $N>2$. Let
$z\equiv\sin^2 (\delta \pi )=0.330992\ldots$ then  $\frac{c}{3-c}=z$ has the unique solution
$N=N_1=\frac{2 \pi }{\arccos \left(\frac{3 z}{z+1}\right)}=8.62247\ldots$.
Hence for all integer values of $N\ge N_0=9$ we have
\begin{eqnarray}\label{LM1a}
 z&<&\frac{c}{3-c}\;,\\
  \label{LM1b}
  0 &<&c-(3-c)z\quad \mbox{  and  } z-1 < 0\;,\\
   \label{LM1c}
   0&>&(z-1)(c-(3-c)z)=z(3+(c-3)z)-c\;,\\
    \label{LM1d}
    c&>&z(3+(c-3)z)\;,\\
   \label{LM1e}
    f(\pi/2)&=& 4(3+c)c>4(3+c)z(3+(c-3)z)=f(x_B)  \;,
\end{eqnarray}
which proves (i). \\
(iv) Let $x_m$ be the unique solution of $f(x_m)=f(\pi/2)$ with $0<x_m<\pi/2$, see Figure \ref{FIGfx}.
Obviously, $x_m < x_1$. By (i) we have $x_B<x_m$. Now consider an arbitrary $x$ with $x_B\le x \le \frac{\pi}{2}$.
If $x\le x_m$ then $x<x_1$ and the claim follows from the increase of $f(x)$ in the interval $[0,x_1]$.
If $x>x_m$ then $f(x)\ge f(\pi/2)>f(x_B)$, the last inequality following from (i). Hence also in this case the claim holds.\\
(iii) Since $x\le x_B <x_1$ the claim follows from the increase of $f(x)$ in the interval $[0,x_1]$. \\
(ii) This follows analogously since $x\le x_A <x_1$. \hfill$\Box$\\

In the following we will use the elementary inequalities
\begin{equation}\label{ineq3}
 |\sin  x|< x \mbox{ and hence } |\csc x|> \frac{1}{x} \mbox{ for } x>0
 \;.
\end{equation}
For a lower bound of the $\sin$ function we will also use
\begin{equation}\label{ineq4}
\sin \pi x > 2\sqrt{2} x
 \mbox{ and  } \csc \pi x< \frac{1 }{ 2\sqrt{2} x } \mbox{ for } 0<x<\frac{1}{4}
 \;.
\end{equation}
Now we can apply lemma \ref{Lemmamono} (ii) and conclude
\begin{eqnarray}
 \nonumber 
 {\mathcal P}^{(\nu)}_{11} =f(x)&\ge& f(x_0)=4\left(3+\cos\frac{2\pi}{N} \right)
 \left(3+\left(\cos\frac{2\pi}{N}-3\right)\sin^2\frac{\pi}{N} \right)\sin^2\frac{\pi}{N} \\
 &&
 \label{LB1a}\\
 \label{LB1b}
 &\ge&
 4\left(3+\cos\frac{2\pi}{N_0} \right)
 \left(3+\left(\cos\frac{2\pi}{N_0}-3\right)\sin^2\frac{\pi}{N_0} \right)\sin^2\frac{\pi}{N}
 \;.\\
 \label{LB1c}
  {\mathcal P}^{(\nu)}_{11}\csc^6 \frac{\pi}{N}
  &\ge&
  4\left(3+\cos\frac{2\pi}{N_0} \right)
 \left(3+\left(\cos\frac{2\pi}{N_0}-3\right)\sin^2\frac{\pi}{N_0} \right)\csc^4 \frac{\pi}{N}\\
 \label{LB1d}
 &\stackrel{(\ref{ineq3})}{>}&
 4\left(3+\cos\frac{2\pi}{N_0} \right)
 \left(3+\left(\cos\frac{2\pi}{N_0}-3\right)\sin^2\frac{\pi}{N_0} \right) \frac{N^4}{\pi^4}\\
  \label{LB1e}
 &\equiv& B_{A1\alpha P1}
 \;.
 \end{eqnarray}
For the inequality (\ref{LB1b}) we have used the increase of the functions $N\mapsto\cos\frac{2\pi}{N}$ and
$N\mapsto -\sin^2\frac{\pi}{N}$ for $N>2$. \\

\begin{figure}
\begin{center}
\includegraphics[clip=on,width=150mm,angle=0]{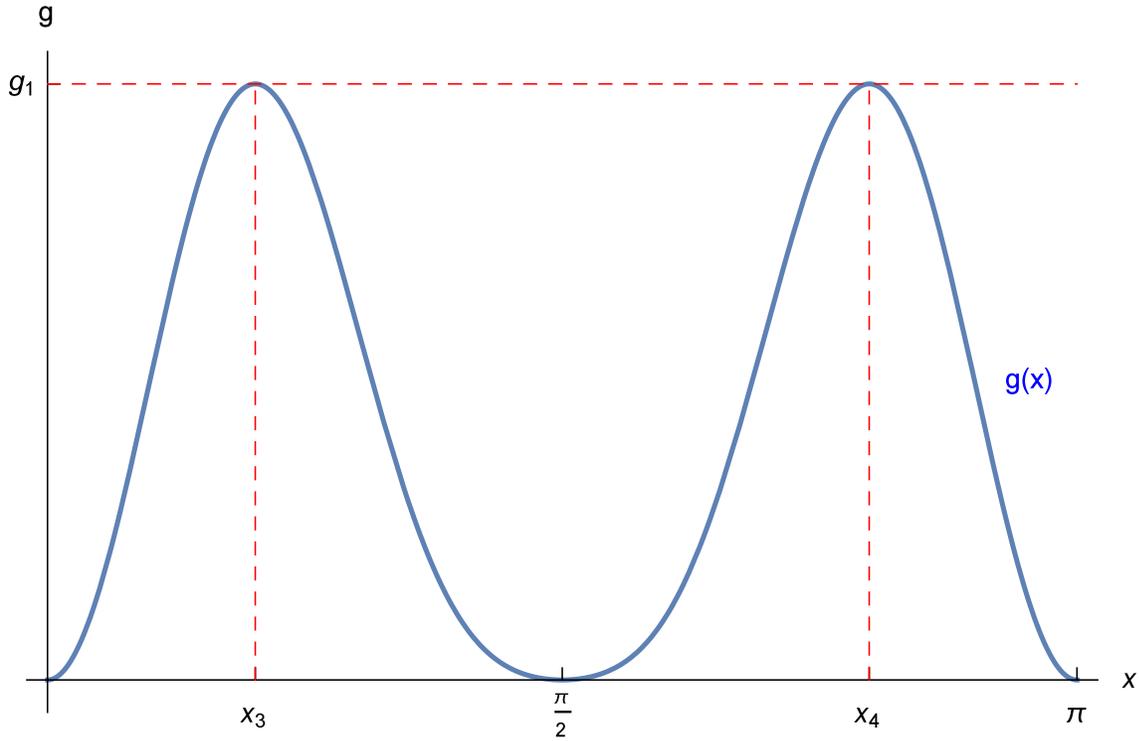}
\end{center}
\caption{Typical form of the function $g(x)$, see (\ref{LB21}) with $N=5$.
\label{FIGgx}
}
\end{figure}

We proceed with the second term at the r.~h.~s.~of (\ref{A1P1}) corresponding to $\lambda=1$ and $\mu=2$.
After some simplifications we conclude, again setting $x=\frac{\pi\nu}{N},\; c=\cos\frac{2\pi}{N}$, and $c_2=\cos\frac{4\pi}{N}$:
\begin{equation}\label{LB21}
 {\mathcal P}^{(\nu)}_{12}=4\left(3+c_2 \right)\left(3+\left(c-3\right)\sin^2 x \right)\sin^2 2x
 \equiv g(x)
 \;.
\end{equation}
A typical graph of the function $g(x)$ is displayed in Figure \ref{FIGgx}. In order to investigate its first maximum we calculate
the derivative and its expansion at $x=0$:
\begin{eqnarray}\nonumber
\frac{d\,g(x)}{dx}&=&32 (c_2+3) \sin 2 x \left(c \sin ^2 x (3 \cos 2 x+1)+(9 \cos 2 x-3) \cos ^2 x\right)\\
   && \label{gsa}\\
   \label{gsb}
   &=& 12\,x+O(x^3)
   \;.
\end{eqnarray}
$\frac{d\,g(x)}{dx}$ is hence positive for sufficiently small values of $x>0$.
To find its first zero $x_3>0$ we introduce the new variable $y=\sin ^2(x)$ and
obtain
\begin{equation}\label{gs2}
 \frac{d\,g(x)}{dx}=64 (c_2+3) \sqrt{(1-y) y} \left(3 (3-c) y^2+2 (c-6) y+3\right)
 \;.
\end{equation}
The last bracket in  (\ref{gs2}) has a unique zero $y_3$ with $0<y_3<1$ of the form
\begin{equation}\label{zero1}
y_3=\frac{\sqrt{c^2-3 c+9}+c-6}{3 (c-3)}
\;.
\end{equation}
The coordinates $x_{3,4}$  of the two maxima of $g(x)$ are the two solutions of the equation
\begin{equation}\label{zero2}
\sin^2 x=\frac{\sqrt{c^2-3 c+9}+c-6}{3 (c-3)}
\end{equation}
in the interval $[0,\pi]$ such that $0<x_3<\pi/2<x_4<\pi$.

We want to show the following
\begin{lemma}\label{Lmono}
$x_3$ is a  increasing function of $N$.
\end{lemma}
Proof: Since the functions $N\mapsto c=\cos\frac{2\pi}{N}$ and $y_3\mapsto x_3=\arcsin \sqrt{y_3}$
are obviously  increasing for $N>2$ it remains to show that $c\mapsto y_3=\frac{\sqrt{c^2-3 c+9}+c-6}{3 (c-3)}$ is
increasing. This follows from
\begin{eqnarray}
 4(c^2-3 c+9)-(3+c)^2  &=& 3 (-3+c)^2 > 0\;, \\
   4(c^2-3 c+9) &>& (3+c)^2 \;, \\
  2\sqrt{c^2-3 c+9} &>&3+c \;,\\
 -3-c+ 2\sqrt{c^2-3 c+9}&>& 0 \;,\\
  \frac{-3-c+ 2\sqrt{c^2-3 c+9}}{2 (-3+c)^2 \sqrt{c^2-3 c+9}}&>&0
  \;,
\end{eqnarray}
and
\begin{equation}\label{mono2}
  \frac{d}{dc}\,\frac{\sqrt{c^2-3 c+9}+c-6}{3 (c-3)}= \frac{-3-c+ 2\sqrt{c^2-3 c+9}}{2 (-3+c)^2 \sqrt{c^2-3 c+9}}
  \;,
\end{equation}
thereby completing the proof of lemma \ref{Lmono}. \hfill$\Box$\\

Recall that the cases A and B are characterized by the inequalities $1\le \nu \le \nu_A=\varepsilon \sqrt{N}$
and $\nu_A < \nu <\nu_B=\delta N$. As above we rewrite these inequalities as
$x_0=\frac{\pi}{N}\le x \le x_A= \frac{\pi\,\varepsilon}{\sqrt{N}}$ and $x_A< x <x_B= \delta \pi$. In order to find a lower bound
for  ${\mathcal P}^{(\nu)}_{12}$ we have to establish the inequality $x_A\le x_3(N)$. Anticipating the analogous problem in the
case B1$\alpha$P we will instead show $x_B\le x_3(N)$, which implies the first inequality since $x_A<x_B$ for
$N>\frac{\varepsilon^2}{\delta^2}=6.32095\ldots$. In view of lemma \ref{Lmono} it suffices to show $x_B\le x_3(N_0)=x_3(9)$ since
$N_0\le N$ implies $x_3(N_0)\le x_3(N)$. The claim then follows from
\begin{eqnarray}\label{mono3a}
\delta&=&\frac{8}{41}=0.195122\ldots<\frac{1}{\pi}\;x_3(9)\\
\label{mono3b}
&=&\frac{1}{\pi}\arcsin \left(\sqrt{\frac{6-\cos \left(\frac{2 \pi
   }{9}\right)-\sqrt{9+\left(-3+\cos \left(\frac{2 \pi }{9}\right)\right)
   \cos \left(\frac{2 \pi }{9}\right)}}{3 \left(3-\cos \left(\frac{2 \pi   }{9}\right)\right)}}\right)\\
   \label{mono3c}
 &  =&0.195737\ldots\;.
\end{eqnarray}
Summarizing, we have shown
\begin{equation}\label{mono4}
x_0\le x\le x_A \mbox{ implies } g(x)\ge g(x_0)
\;,
\end{equation}
and hence
\begin{eqnarray}\label{mono5a}
  {\mathcal P}^{(\nu)}_{1,2} &=& g(x)\ge g(x_0) \\
  \label{mono5b}
   &=& \sin ^2\frac{2 \pi }{N} \left(8 \cos\frac{2 \pi   }{N}
   -\cos\frac{4 \pi }{N}+5\right) \left(\cos  \frac{4 \pi }{N}+3\right) \\
   \label{mono5c}
   &\ge&\sin ^2\frac{2 \pi }{N} \left(8 \cos \frac{2 \pi   }{N_0}
   -\cos\frac{4 \pi }{N_0}+5\right) \left(\cos  \frac{4 \pi }{N_0}+3\right)
  .
\end{eqnarray}
The last inequality uses the monotonic increase of the function
$N\mapsto \left(8 \cos \frac{2 \pi   }{N}
   -\cos \frac{4 \pi }{N}+5\right) \left(\cos
  \frac{4 \pi }{N}+3\right)$. For the last bracket this is obvious; for the first bracket
it follows from $\left(8 \cos\frac{2 \pi   }{N}   -\cos \frac{4 \pi }{N}+5\right)
=6 + 8 c - 2 c^2$ and $\frac{d}{dc}\left(6 + 8 c - 2 c^2\right)=8-4c>0$ for $|c|<1$.
Further, using
\begin{equation}\label{mono6}
\sin ^2\frac{2 \pi }{N}= 4 \,\sin^2\frac{ \pi }{N}\,\cos^2\frac{ \pi }{N}
\ge 4 \,\sin^2\frac{ \pi }{N}\,\cos^2\frac{ \pi }{N_0}
\end{equation}
we obtain from (\ref{mono5c}):
\begin{eqnarray}\label{LB2a}
  && {\mathcal P}^{(\nu)}_{1,2}\csc^3\frac{\pi}{N}\csc^3\frac{2\pi}{N} \\
  \label{LB2b}
 &\ge& 4 \cos^2\frac{\pi}{N_0} \left(8 \cos \frac{2 \pi   }{N_0}
   -\cos\frac{4 \pi }{N_0}+5\right) \left(\cos  \frac{4 \pi }{N_0}+3\right)
   \csc\frac{\pi}{N}\csc^3\frac{2\pi}{N}\\
   \label{LB2c}
   &\stackrel{(\ref{ineq3})}{>}& 4 \cos^2\frac{\pi}{N_0}
   \left(8 \cos \frac{2 \pi   }{N_0}
   -\cos\frac{4 \pi }{N_0}+5\right) \left(\cos  \frac{4 \pi }{N_0}+3\right)\,
 \frac{N}{\pi}\,\frac{N^3}{(2\pi)^3}\\
 \label{LB2d}
 &=&\frac{1}{2\pi^4}\cos^2\frac{\pi}{N_0} \left(8 \cos \frac{2 \pi   }{N_0}
   -\cos\frac{4 \pi }{N_0}+5\right) \left(\cos  \frac{4 \pi }{N_0}+3\right)\,
 N^4\\
 \label{LB2e}
 &\equiv& B_{A1\alpha P2}
 \;.
\end{eqnarray}
Summarizing the equations (\ref{A1P1}), (\ref{LB1e}) and (\ref{LB2e}), we have established the lower bound
\begin{equation}\label{LB3}
 K_{A1\alpha P}>B_{A1\alpha P1}+B_{A1\alpha P2}\equiv B_{A1\alpha P}
 \;.
\end{equation}
Note that both terms $B_{A1\alpha P1}$ and $B_{A1\alpha P2}$ are of order $O(N^4)$.
This completes the case A1$\alpha$P.\\


\fbox{Case A1$\alpha$N}\\

Recall that we are looking for an upper bound of the absolute value of the contribution of all
(possibly) negative terms in the double sum (\ref{Adouble}).
For the present case the partial sum of these terms is
\begin{eqnarray}\label{A1N1a}
  K_{A1\alpha N} &\equiv& \sum_{\lambda=1}^{N/4}  \sum_{\mu=1}^{N/4}{\mathcal N}^{(\nu)}_{\lambda\mu}
  \csc^3\frac{\pi\lambda}{N} \csc^3\frac{\pi\mu}{N}\\
 \label{A1N1b}
 &\stackrel{(\ref{split1c})}{=}&- \sum_{\lambda\mu}\sin\frac{2\pi\lambda}{N}\sin\frac{2\pi\mu}{N}\sin\frac{2\pi\lambda\nu}{N}\sin\frac{2\pi\mu\nu}{N}
 \csc^3\frac{\pi\lambda}{N}\csc^3\frac{\pi\mu}{N}\\ \nonumber
&=&- \sum_{\lambda\mu}\left(2\sin\frac{\pi\lambda}{N}\cos\frac{\pi\lambda}{N}\right)
\left(2\sin\frac{\pi\mu}{N}\cos\frac{\pi\mu}{N}\right)  \\ \label{A1N1c}
 &&\sin\frac{2\pi\lambda\nu}{N}\sin\frac{2\pi\mu\nu}{N}
 \csc^3\frac{\pi\lambda}{N}\csc^3\frac{\pi\mu}{N}\\
 \label{A1N1d}
 &=&-4 \sum_{\lambda\mu}\cos\frac{\pi\lambda}{N}\cos\frac{\pi\mu}{N}\sin\frac{2\pi\lambda\nu}{N}\sin\frac{2\pi\mu\nu}{N}
 \csc^2\frac{\pi\lambda}{N}\csc^2\frac{\pi\mu}{N}
 \;.
\end{eqnarray}
Applying the triangle inequality to the sum (\ref{A1N1d}) we obtain
\begin{eqnarray}\label{A1N2a}
  | K_{A1\alpha N} | &\le& 4 \sum_{\lambda\mu}\left|\sin\frac{2\pi\lambda\nu}{N}\sin\frac{2\pi\mu\nu}{N}\right|
  \csc^2\frac{\pi\lambda}{N}\csc^2\frac{\pi\mu}{N}\\
  \label{A1N2b}
  &\stackrel{(\ref{ineq3})}{<}& 4\sum_{\lambda\mu}\frac{4\pi^2
\,\lambda\,\mu\,\nu^2}{N^2}\csc^2\frac{\pi\lambda}{N}\csc^2\frac{\pi\mu}{N}\\
\label{A1N2c}
  &\stackrel{(\ref{ineq4})}{<}& 4\sum_{\lambda\mu}\frac{4\pi^2
\,\lambda\,\mu\,\nu^2}{N^2}\frac{N^2}{8\lambda^2}\,\frac{N^2}{8\mu^2}\\
  \label{A1N2d}
  &=&\frac{\pi^2}{4}N^2\,\nu^2\,\left(\sum_{\lambda=1}^{N/4}\frac{1}{\lambda} \right)
\,\left(\sum_{\mu=1}^{N/4}\frac{1}{\mu} \right)\\
\label{A1N2e}
&<& \frac{\pi^2}{4}N^3 \varepsilon^2 \left( \log \frac{N}{4}+\gamma+\frac{2}{N}\right)^2\\
\label{A1N2f}
&\le& \frac{\pi^2}{4}N^3 \varepsilon^2 \left( \log \frac{N}{4}+\gamma+\frac{2}{N_0}\right)^2
\equiv B_{A1\alpha N}
\;.
\end{eqnarray}
The  inequality in (\ref{A1N2e}) deserves some explanation. First, we used $\nu\le \varepsilon\sqrt{N}$ according to
the definition of case A in table \ref{Tab2}. Secondly, if $4$ divides $N$ the sums in (\ref{A1N2d}) are the harmonic numbers $H_{N/4}$
and the upper bound involving $\log\frac{N}{4}+\gamma +\frac{2}{N}$ is a standard result, see \cite{H03} pp.~73--75, where $\gamma=0.5772\ldots$
denotes Euler's constant. If $4$ does not divide $N$, the sums in (\ref{A1N2d}) run only up to $\lfloor \frac{N}{4} \rfloor$
and (\ref{A1N2e}) follows from the monotonic increase of the function $x\mapsto \log x +\frac{1}{2x}$ for $x>\frac{1}{2}$.\\

The upper bound  (\ref{A1N2f}) is of order $O(N^3 (\log N)^2)$ which is close to the order $O(N^4)$ of the lower bound
in the case A1$\alpha$P but strictly less, as it must be for the present proof strategy.
Note that without the restriction to $\nu\le \varepsilon\sqrt{N}$
we would not have achieved this result which explains the introduction of the case distinction according to the cases A and B.\\

\fbox{Case A1$\beta$N}\\

We consider
\begin{equation}\label{A1N3}
  K_{A1\beta N} \equiv \sum_{\lambda=1}^{N/4}  \sum_{\mu=N/4}^{N/2}{\mathcal N}^{(\nu)}_{\lambda\mu}
  \csc^3\frac{\pi\lambda}{N} \csc^3\frac{\pi\mu}{N}
  \;.
 \end{equation}

The calculations are analogous to the case A1$\alpha$N except that we now use the estimate
\begin{equation}\label{A1N4}
 \csc^3 \frac{\pi\mu}{N}< 2 \sqrt{2}
 \;,
\end{equation}
that holds since $\sin x>\frac{\sqrt{2}}{2}$ for $\frac{\pi}{4}<x<\frac{3\pi}{4}$. It follows that
\begin{eqnarray}\label{A1N5a}
 \left| K_{A1\beta N}\right| &<& 4 \sqrt{2}\,\sum_{\lambda\mu} \left|\sin\frac{2\pi\lambda\nu}{N}\right| \csc^2\frac{\pi\lambda}{N}\\
 \label{A1N5b}
 &\stackrel{(\ref{ineq3})(\ref{ineq4})}{\le}&4 \sqrt{2}\,\frac{N}{4}\,\sum_{\lambda=1}^{N/4}\frac{2\pi\lambda\nu}{N}\frac{N^2}{8\lambda^2}\\
  \label{A1N5c}
  &=& \frac{\sqrt{2}\pi}{4}\,N^2\,\nu\,\sum_{\lambda=1}^{N/4}\frac{1}{\lambda}\\
  \label{A1N5d}
  &<& \frac{\sqrt{2}\pi\varepsilon}{4}\,N^{3/2}\,\left( \log\frac{N}{4}+\gamma +\frac{2}{N_0}\right)\equiv B_{A1\beta N}
  \;.
\end{eqnarray}
In the inequality (\ref{A1N5b}) we have used the fact that the number of terms in the sum $\sum_{\mu=N/4}^{N/2}\ldots$ does not exceed $N/4$ due
to our convention concerning summation limits.\\

\fbox{Case A2$\alpha$N}\\

We consider the terms according to (\ref{split2c}) and (\ref{split2d}) separately.
\begin{eqnarray}\label{A2N1a}
  \left|K_{A2\alpha N1}\right| &\le& \sum_{\lambda=N/4}^{N/2}\sum_{\mu=1}^{N/4}\left|{\mathcal N}^{(\nu)}_{1,\lambda\mu}\right|
  \csc^3\frac{\pi\lambda}{N} \csc^3\frac{\pi\mu}{N}\\
 \nonumber
 &=& \sum_{\lambda\mu}
\left|\left(3+\cos\frac{2\pi\mu}{N}\right)\cos\frac{2\pi\lambda}{N}\left(1-\cos\frac{2\pi\lambda\nu}{N} \right)\right|\\
 \label{A2N1b}
&&\left(1-\cos\frac{2\pi\mu\nu}{N}\right) \csc^3\frac{\pi\lambda}{N} \csc^3\frac{\pi\mu}{N}\\
\label{A2N1c}
&\stackrel{(\ref{A1N4})}{\le}&  \sum_{\lambda\mu} 8\cdot 2\sin^2\frac{\pi\mu\nu}{N}\; 2\sqrt{2}\; \csc^3\frac{\pi\mu}{N}\\
\label{A2N1d}
 &\stackrel{(\ref{ineq3})(\ref{ineq4})}{\le}&32\sqrt{2}\, \frac{N}{4}\,\sum_{\mu=1}^{N/4}\frac{\pi^2\mu^2\nu^2}{N^2}
  \frac{N^3}{(2\sqrt{2}\mu)^3} \\
 \label{A2N1e}
 &=& \frac{\pi^2}{2}\,N^2\,\nu^2\,\sum_{\mu=1}^{N/4} \frac{1}{\mu}\\
  \label{A2N1f}
&\le& \frac{\pi^2\varepsilon^2}{2}\,N^3\,\left( \log\frac{N}{4}+\gamma +\frac{2}{N_0}\right)\equiv B_{A2\alpha N1}
  \;.
\end{eqnarray}
In the inequality (\ref{A2N1c}) we have used the obvious bound
$\left| \left(3+\cos\frac{2\pi\mu}{N}\right)\cos\frac{2\pi\lambda}{N}\left(1-\cos\frac{2\pi\lambda\nu}{N} \right)\right|\le 4\cdot 1\cdot 2=8$.

\begin{eqnarray}\label{A2N2a}
  \left|K_{A2\alpha N2}\right| &\le& \sum_{\lambda=N/4}^{N/2}\sum_{\mu=1}^{N/4}\left|{\mathcal N}^{(\nu)}_{2,\lambda\mu}\right|
  \csc^3\frac{\pi\lambda}{N} \csc^3\frac{\pi\mu}{N}\\
 \nonumber
 &=& \sum_{\lambda\mu}
\left|\sin\frac{2\pi\lambda}{N}\sin\frac{2\pi\mu}{N}\sin\frac{2\pi\lambda\nu}{N}\sin\frac{2\pi\mu\nu}{N}\right|\\
 \label{A2N2b}
&& \csc^3\frac{\pi\lambda}{N} \csc^3\frac{\pi\mu}{N}\\
\label{A2N2c}
&\stackrel{(\ref{A1N4})}{\le}& \sum_{\lambda\mu}\left( 2\sin\frac{\pi\mu}{N}\cos\frac{\pi\mu}{N}\right) \left|\sin\frac{2\pi\mu\nu}{N}\right|
\; 2\sqrt{2}\; \csc^3\frac{\pi\mu}{N}\\
\label{A2N2d}
 &\stackrel{(\ref{ineq3})}{\le}&4\sqrt{2}\, \frac{N}{4}\,\sum_{\mu=1}^{N/4}\frac{2\pi\mu\nu}{N}\,\csc^2\frac{\pi\mu}{N }
   \\
 \label{A2N2e}
 &\stackrel{(\ref{ineq4})}{\le}&2\sqrt{2}\pi\,\nu\,\sum_{\mu=1}^{N/4} \mu\,\frac{N^2}{8\mu^2}\\
  \label{A2N2f}
&<& \frac{\sqrt{2}\pi\varepsilon}{4}\,N^{5/2}\,\left( \log\frac{N}{4}+\gamma +\frac{2}{N_0}\right)\equiv B_{A2\alpha N1}
  \;.
\end{eqnarray}

\fbox{Case A2$\beta$N}\\

We obtain
\begin{eqnarray}\label{A2N3a}
  \left|K_{A2\beta N}\right| &\le& \sum_{\lambda=N/4}^{N/2}\sum_{\mu=N/4}^{N/2}
  \left|{\mathcal N}^{(\nu)}_{1,\lambda\mu}+{\mathcal N}^{(\nu)}_{2,\lambda\mu}\right|
  \csc^3\frac{\pi\lambda}{N} \csc^3\frac{\pi\mu}{N}\\
 \nonumber
 &=& \sum_{\lambda\mu}
 \left|\left(3+\cos\frac{2\pi\mu}{N}\right)\cos\frac{2\pi\lambda}{N}\left(1-\cos\frac{2\pi\lambda\nu}{N} \right)\right.\\
 \nonumber
&&\left. \left(1-\cos\frac{2\pi\mu\nu}{N}\right) -
\sin\frac{2\pi\lambda}{N}\sin\frac{2\pi\mu}{N}\sin\frac{2\pi\lambda\nu}{N}\sin\frac{2\pi\mu\nu}{N}\right|\\
 \label{A2N3b}
&& \csc^3\frac{\pi\lambda}{N} \csc^3\frac{\pi\mu}{N}\\
\label{A2N3c}
&\stackrel{(\ref{A1N4})}{\le}&  13\cdot \left(\frac{N}{4}\right)^2 \left(\sqrt{2}\right)^6=\frac{13}{2}\,N^2\equiv B_{A2\beta N1}
  \;.
\end{eqnarray}
In the inequality (\ref{A2N3c}) we have used that for $N/4\le \mu \le N/2$ we have $\cos\frac{2\pi\mu}{N}\le 0$ and hence the terms
$\left|\ldots \right|$ can be bounded by $3\cdot 1\cdot 2\cdot 2 +1 =13$.\\

Now we turn to the case B defined by $\varepsilon\,\sqrt{N}<\nu<\delta N$.\\

\fbox{Case B1$\alpha$P}\\

The calculations are similar to the case A1$\alpha$P.
\begin{equation}\label{B1P1}
 K_{B1\alpha P}\equiv \sum_{\lambda=1}^{N/4}\sum_{\mu=1}^{N/4}
 {\mathcal P}^{(\nu)}_{\lambda\mu} \csc^3 \frac{\pi\lambda}{N}\, \csc^3 \frac{\pi\mu}{N}
 >  {\mathcal P}^{(\nu)}_{11} \csc^6 \frac{\pi}{N}+ {\mathcal P}^{(\nu)}_{12} \csc^3 \frac{\pi}{N}\csc^3 \frac{2\pi}{N}
 \;.
\end{equation}
Let us begin with the first term at the r.~h.~s.~of (\ref{B1P1}) corresponding to $\lambda=\mu=1$.
Here we can rely on the result of lemma \ref{Lmono} (iii) in order to find a lower bound of ${\mathcal P}^{(\nu)}_{11}$:
\begin{eqnarray}\label{B1P2a}
  {\mathcal P}^{(\nu)}_{11}&=&f(x)\ge f(x_A)\\ \label{B1P2b}
  &=& 4\left(3+\cos\frac{2\pi}{N}\right)
  \left( 3+\left( \cos\frac{2\pi}{N}-3\right)\sin^2\frac{\pi\varepsilon}{\sqrt{N}}\right)\sin^2\frac{\pi\varepsilon}{\sqrt{N}}\\ \label{B1P2c}
&\ge& 4\left(3+\cos\frac{2\pi}{N_0}\right)
  \left( 3+\left( \cos\frac{2\pi}{N_0}-3\right)\sin^2\frac{\pi\varepsilon}{\sqrt{N_0}}\right)\sin^2\frac{\pi\varepsilon}{\sqrt{N}}\\
  \label{B1P2d}
  &\stackrel{(\ref{ineq4})}{\ge}&
  4\left(3+\cos\frac{2\pi}{N_0}\right)
  \left( 3+\left( \cos\frac{2\pi}{N_0}-3\right)\sin^2\frac{\pi\varepsilon}{\sqrt{N_0}}\right)
  \frac{8\varepsilon^2}{N}\;,
\end{eqnarray}
where we have used $N>16 \varepsilon^2=3.85048\ldots$ and hence $\frac{\varepsilon}{\sqrt{N}}=x<\frac{1}{4}$  in order to apply (\ref{ineq4}) in the last
inequality (\ref{B1P2d}). Further we conclude
\begin{equation}\label{B1P3}
 {\mathcal P}^{(\nu)}_{11}\csc^6 \frac{\pi}{N}\stackrel{(\ref{ineq3})}{>}
\frac{32 \varepsilon^2}{\pi^6}N^5
\left(3+\cos\frac{2\pi}{N_0}\right)
  \left( 3+\left( \cos\frac{2\pi}{N_0}-3\right)\sin^2\frac{\pi\varepsilon}{\sqrt{N_0}}\right)
  \equiv B_{B1\alpha P1}
  \;.
\end{equation}

We proceed with the second term at the r.~h.~s.~of (\ref{B1P1}) corresponding to $\lambda=1,\;\mu=2$. According to the results
derived in the case A1$\alpha$P we conclude
\begin{eqnarray}\label{B1P4a}
  {\mathcal P}^{(\nu)}_{12}&=&g(x)\ge g(x_A)\\ \nonumber
  &=& 2^4\left(3+\cos\frac{4\pi}{N}\right)
  \left( 3+\left( \cos\frac{2\pi}{N}-3\right)\sin^2\frac{\pi\varepsilon}{\sqrt{N}}\right)\cos^2\frac{\pi\varepsilon}{\sqrt{N}}
  \sin^2\frac{\pi\varepsilon}{\sqrt{N}}\\
  &&\label{B1P4b}\\
  \nonumber
&\ge& 2^4\left(3+\cos\frac{4\pi}{N_0}\right)
  \left( 3+\left( \cos\frac{2\pi}{N_0}-3\right)\sin^2\frac{\pi\varepsilon}{\sqrt{N_0}}\right)\cos^2\frac{\pi\varepsilon}{\sqrt{N_0}}
  \sin^2\frac{\pi\varepsilon}{\sqrt{N}}\\
  &&\label{B1P4c}\\
  &\stackrel{(\ref{ineq4})}{>}&
  2^7\left(3+\cos\frac{4\pi}{N_0}\right)
  \left( 3+\left( \cos\frac{2\pi}{N_0}-3\right)\sin^2\frac{\pi\varepsilon}{\sqrt{N_0}}\right)\cos^2\frac{\pi\varepsilon}{\sqrt{N_0}}
  \frac{\varepsilon^2}{N}
  \;.
\end{eqnarray}
Hence
\begin{eqnarray}\nonumber
&& {\mathcal P}^{(\nu)}_{12}\csc^3 \frac{\pi}{N}\csc^3 \frac{2\pi}{N}\\
\label{B1P5a}
& \stackrel{(\ref{ineq3})}{>}&
\frac{16 \varepsilon^2}{\pi^6}N^5
\left(3+\cos\frac{4\pi}{N_0}\right)
  \left( 3+\left( \cos\frac{2\pi}{N_0}-3\right)\sin^2\frac{\pi\varepsilon}{\sqrt{N_0}}\right)\cos^2\frac{\pi\varepsilon}{\sqrt{N_0}}\\
 \label{B1P5b}
  &\equiv& B_{B1\alpha P2}
  \;.
\end{eqnarray}

Summarizing the equations (\ref{B1P1}), (\ref{B1P3}) and (\ref{B1P5b}), we have established the lower bound
\begin{equation}\label{B1P6}
 K_{B1\alpha P}>B_{B1\alpha P1}+B_{B1\alpha P2}\equiv B_{B1\alpha P}
 \;.
\end{equation}

Next we consider terms of the double sum (\ref{Adouble}) that are possibly negative. Since we need not make any assumption
about the range of $\nu$ the results are valid for both cases, B and C. It turns out that the part of  ${\mathcal N}^{(\nu)}_{\lambda\mu}$
that contains only $\sin-$ terms can be treated separately without making the case distinctions due to $1,2$ and $\alpha,\beta$.
We will call this the case BNs.\\

\fbox{Case BNs$=$CNs}\\

We will extend the summations to the whole domain $1\le\lambda\le N-1$ and $1\le\mu\le N-1$ and accordingly introduce a factor $\frac{1}{4}$
in order to comply with our convention explained above.
\begin{eqnarray}\label{B1N1a}
  K_{B N s} &\equiv&\frac{1}{4} \sum_{\lambda=1}^{N-1}  \sum_{\mu=1}^{N-1}{\mathcal N}^{(\nu)}_{2,\lambda\mu}
  \csc^3\frac{\pi\lambda}{N} \csc^3\frac{\pi\mu}{N}\\
 \label{B1N1b}
 &\stackrel{(\ref{split1c})}{=}&-\frac{1}{4} \sum_{\lambda\mu}\sin\frac{2\pi\lambda}{N}\sin\frac{2\pi\mu}{N}
 \sin\frac{2\pi\lambda\nu}{N}\sin\frac{2\pi\mu\nu}{N}
 \csc^3\frac{\pi\lambda}{N}\csc^3\frac{\pi\mu}{N}\\ \nonumber
&=&- \frac{1}{4}\sum_{\lambda\mu}\left(2\sin\frac{\pi\lambda}{N}\cos\frac{\pi\lambda}{N}\right)
\left(2\sin\frac{\pi\mu}{N}\cos\frac{\pi\mu}{N}\right)  \\ \label{B1N1c}
 &&\sin\frac{2\pi\lambda\nu}{N}\sin\frac{2\pi\mu\nu}{N}
 \csc^3\frac{\pi\lambda}{N}\csc^3\frac{\pi\mu}{N}\\
 \label{B1N1d}
 &=&- \sum_{\lambda\mu}\cos\frac{\pi\lambda}{N}\cos\frac{\pi\mu}{N}\sin\frac{2\pi\lambda\nu}{N}\sin\frac{2\pi\mu\nu}{N}
 \csc^2\frac{\pi\lambda}{N}\csc^2\frac{\pi\mu}{N}
 \;.
\end{eqnarray}

Hence
\begin{eqnarray}\label{B1N2a}
  \left|K_{B Ns}\right| &\le& \sum_{\lambda\mu}\csc^2\frac{\pi\lambda}{N}\csc^2\frac{\pi\mu}{N}\\
\label{B1N2b}
&=& \left( \frac{N^2-1}{3}\right)^2
\equiv B_{BNs}\equiv B_{CNs}
 \;.
\end{eqnarray}
For equation (\ref{B1N2b}) we have used \cite{H75}, (24.1.2).\\

\fbox{Case B2$ \alpha$N $=$ C2$\alpha$N}\\

We consider only the terms according to (\ref{split2c}) since the other terms are already included in the case BNs$=$CNs.
\begin{eqnarray}\label{B2N1a}
  K_{B2\alpha N1}&\equiv& \sum_{\lambda=N/4}^{N/2}\sum_{\mu=1}^{N/4}
 {\mathcal N}^{(\nu)}_{1,\lambda\mu} \csc^3 \frac{\pi\lambda}{N}\, \csc^3 \frac{\pi\mu}{N}\;,\\
 \label{B2N1b}
 \left| K_{B2\alpha N1}\right|&\le& \sum_{\lambda\mu}\left|{\mathcal N}^{(\nu)}_{1,\lambda\mu}\right|
 \csc^3 \frac{\pi\lambda}{N}\, \csc^3 \frac{\pi\mu}{N}\\
 \nonumber
 &=&\sum_{\lambda\mu}\left|\left( 3+\cos\frac{2\pi\mu}{N}\right)\cos\frac{2\pi\lambda}{N}\left(1-\cos\frac{2\pi\lambda\nu}{N}\right)\right|\\
 \label{B2N1c}
&& \left|1-\cos\frac{2\pi\mu\nu}{N}\right| \csc^3 \frac{\pi\lambda}{N}\, \csc^3 \frac{\pi\mu}{N} \\
 \label{B2N1d}
&\stackrel{(\ref{A1N4})}{\le}& 16\cdot 2\sqrt{2}\,\frac{N}{4}\sum_{\mu=1}^{N/4}\csc^3 \frac{\pi\mu}{N} \\
\label{B2N1e}
&\stackrel{(\ref{ineq4})}{<}&8\, \sqrt{2}\,N\sum_{\mu=1}^{\infty}\left( \frac{N}{2\sqrt{2}\mu}\right)^3 \\
\label{B2N1f}
&=& \frac{1}{2}N^4\zeta(3)\equiv B_{B2\alpha N1}\equiv B_{C2\alpha N1}
\;.
\end{eqnarray}
\\

\fbox{Case B2$\beta$N $=$ C2$\beta$N}\\

\begin{eqnarray}\label{B2N3a}
  K_{B2\beta N}&\equiv& \sum_{\lambda=N/4}^{N/2}\sum_{\mu=N/4}^{N/2}
 {\mathcal N}^{(\nu)}_{1,\lambda\mu} \csc^3 \frac{\pi\lambda}{N}\, \csc^3 \frac{\pi\mu}{N}\;,\\
 \label{B2N3b}
 \left| K_{B2\beta N}\right|&\le& \sum_{\lambda\mu}\left|{\mathcal N}^{(\nu)}_{1,\lambda\mu}\right|
 \csc^3 \frac{\pi\lambda}{N}\, \csc^3 \frac{\pi\mu}{N}\\
 \nonumber
 &=&\sum_{\lambda\mu}\left|\left( 3+\cos\frac{2\pi\mu}{N}\right)\cos\frac{2\pi\lambda}{N}\left(1-\cos\frac{2\pi\lambda\nu}{N}\right)
 \left(1-\cos\frac{2\pi\mu\nu}{N}\right)\right|\\
 \label{B2N3c}
&&
\csc^3 \frac{\pi\lambda}{N}\, \csc^3 \frac{\pi\mu}{N} \\
 \label{B2N3d}
&\stackrel{(\ref{A1N4})}{\le}& 12\cdot \left(2\sqrt{2}\right)^2\,\left(\frac{N}{4}\right)^2 = 6\,N^2\equiv B_{B2\beta N}\equiv B_{C2\beta N}
\;.
\end{eqnarray}
In the inequality (\ref{B2N3d}) we have used that for $N/4 \le \mu\le N/2$ we have $\cos\frac{2\pi\mu}{N}\le 0$ and hence the term $\left|\ldots\right|$
can be bounded by $3\cdot 1\cdot 2\cdot2 =12$.\\

\begin{table}
\caption{{\label{Tab3}}Table of the leading order w.~r.~t.~$N$ of the bounds for the various cases according to table \ref{Tab2}.\\}
\begin{center}\begin{tabular}{||c|c||c|c||c|c||c|c||}\hline\hline
Case & Order&$\ldots$&&&&&\\
\hline\hline
 A1$\alpha$P & $N^4$&&&&&&\\ \hline
 A1$\alpha$N& $N^3 (\log N)^2$ & A1$\beta$N & $N^{3/2} \log N$ &  A2$\alpha$N & $N^{5/2} \log N $  &  A2$\beta$N & $N^2 $ \\
 \hline\hline
  B1$\alpha$P & $N^5$&&&&&&\\ \hline
 BNs& $N^4$  &  B2$\alpha$N & $N^4$  &  B2$\beta$N & $N^2$ && \\
 \hline\hline
  C1$\alpha$P & $N^6$&&&&&&\\ \hline
 CNs& $N^4$ &  C2$\alpha$N & $N^4$  &  C2$\beta$N & $N^2 $ &&\\
 \hline\hline
\end{tabular}\end{center}
\end{table}

Now we consider the positive terms of the double sum (\ref{Adouble}) in the case C defined by $\delta N\le\nu\le \frac{N}{2}$ and .\\

\fbox{Case C1$\alpha$P}\\

The calculations are similar to the case A1$\alpha$P except that we only consider one term of the double sum (\ref{Adouble}) as a lower bound.
\begin{equation}\label{C1P1}
 K_{C1\alpha P}\equiv \sum_{\lambda=1}^{N/4}\sum_{\mu=1}^{N/4}
 {\mathcal P}^{(\nu)}_{\lambda\mu} \csc^3 \frac{\pi\lambda}{N}\, \csc^3 \frac{\pi\mu}{N}
 >  {\mathcal P}^{(\nu)}_{11} \csc^6 \frac{\pi}{N}
 \;.
\end{equation}

Again we utilize the result of lemma \ref{Lmono} (iv) in order to find a lower bound of ${\mathcal P}^{(\nu)}_{11}$:
\begin{eqnarray}\label{C1P2a}
  {\mathcal P}^{(\nu)}_{11}&=&f(x)\ge f(x_B)\\ \label{C1P2b}
  &=& 4\left(3+\cos\frac{2\pi}{N}\right)
  \left( 3+\left( \cos\frac{2\pi}{N}-3\right)\sin^2\delta \pi\right)\sin^2\delta \pi\\
   \label{C1P2c}
&\ge& 4\left(3+\cos\frac{2\pi}{N_0}\right)
  \left( 3+\left( \cos\frac{2\pi}{N_0}-3\right)\sin^2\delta \pi\right)\sin^2\delta \pi
  \;.
\end{eqnarray}
Hence we conclude
\begin{equation}\label{C1P3}
 {\mathcal P}^{(\nu)}_{11} \csc^6 \frac{\pi}{N}\stackrel{(\ref{ineq3})}{>}
\frac{4}{\pi^6}\,N^6\,
\left(3+\cos\frac{2\pi}{N_0}\right)
  \left( 3+\left( \cos\frac{2\pi}{N_0}-3\right)\sin^2\delta \pi\right)\sin^2\delta \pi
  \equiv B_{C1\alpha P}
  \;.
\end{equation}

\begin{figure}
\begin{center}
\includegraphics[clip=on,width=150mm,angle=0]{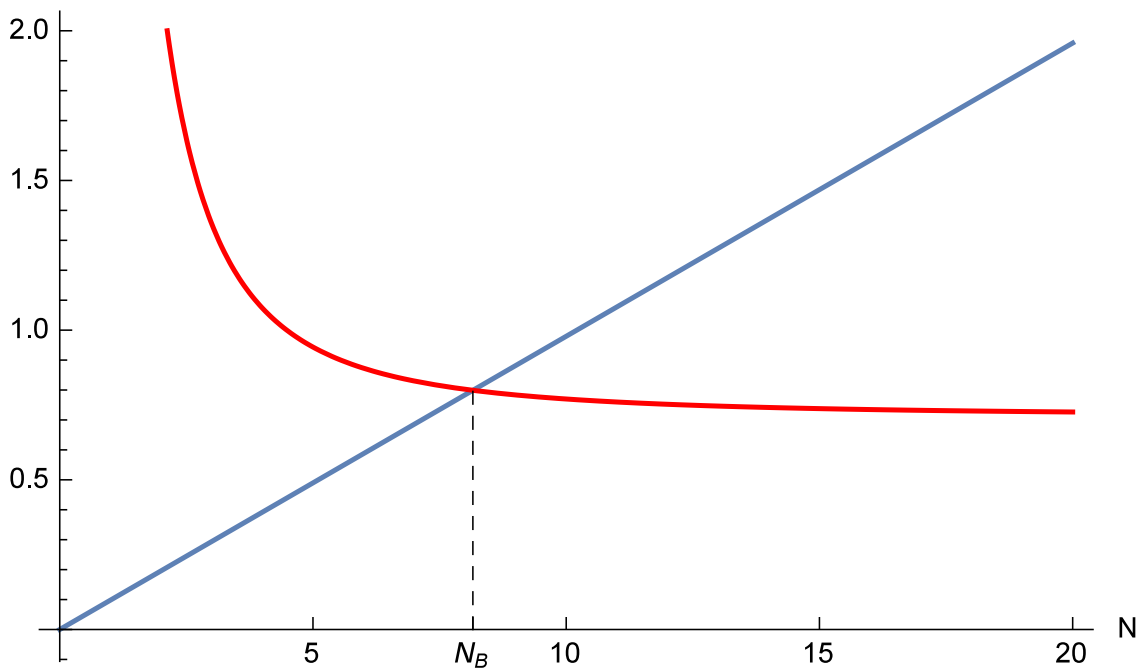}
\end{center}
\caption{Intersection of the scaled bounds  $N^{-4}B_{1\alpha P}$ (blue graph) and $ N^{-4}B_{NB}$ (red graph) at $N_B=8.15728\ldots$ .
\label{FIGib}
}
\end{figure}

The complete results are displayed in table \ref{Tab3} as far as the leading order w.~r.~t.~$N$ is concerned. We note that for each case,
A, B or C, the leading order (of the lower bound) of the positive terms is larger than the leading order (of the upper bound of the absolute value)
of the possibly negative terms. This implies that $\det K^{(\nu)}>0$ for all $\nu=1,\ldots N-1$ and sufficiently large $N$. Hence the following holds:
\begin{theorem}\label{TGS}
  There exists an $N_\ast$ such that for all $N\ge N_\ast$ the state $\pm {\mathbf t}$,
  see (\ref{R4}), is a ground state of the dipole ring of length $N$.
\end{theorem}
In the remainder of this section we will show that the number $N_\ast$ in theorem \ref{TGS} can be chosen as $N_\ast=9$
which is compatible with the assumption $N\ge N_0=9$ we have made from the outset, see (\ref{ngn0}).
We begin with the cases B and C that turn out to be simpler than the case A. \\

\fbox{Case B}\\

After inserting (\ref{epsilon}), (\ref{delta})
and $N_0=9$ we divide the lower bound of the positive terms (\ref{B1P6}) by $N^4$ and obtain
\begin{eqnarray}
 \nonumber 
  N^{-4}B_{B1\alpha P}&=& \frac{10816}{2809 \,\pi ^6}N \left(6+2 \cos \frac{2 \pi }{9}+\left(3+\sin\frac{\pi }{18}\right) \cos^2\frac{26 \pi }{159}\right)
  \\
  \label{Bound1a}
  &&
    \left(3+\sin ^2\frac{26\pi }{159} \left(\cos\frac{2 \pi}{9}-3\right)\right)=0.0979486\ldots N
    \;.
 \end{eqnarray}
This is an increasing linear function of $N$. On the other hand, the upper bound of the absolute value of the possibly negative terms
is evaluated as follows:
\begin{eqnarray}\label{Bound2a}
   N^{-4}B_{NB}&\equiv& N^{-4}\left(B_{BNs}+ B_{B2\alpha N1}+B_{B2\beta N} \right) \\
   \label{Bound2b}
   &=&\frac{1}{18} \left(\frac{2}{N^4}+\frac{104}{N^2}+9 \zeta(3)+2\right)
   \;.
  \end{eqnarray}
  This is a decreasing function of $N>0$ with the limits $\frac{1}{18} \left(9\, \zeta(3)+2\right)=0.71214\ldots$
   for $N\longrightarrow\infty$ and $\infty$ for $N\longrightarrow 0$. The graphs of both functions (\ref{Bound1a}) and (\ref{Bound2b}) hence intersect at a unique point with
  $N_B=8.15728\ldots$, see Figure \ref{FIGib}. Hence for $N\ge N_\ast=9 >N_B$ we have $B_{B1\alpha P}>B_{NB}$ and
  the state $\pm {\mathbf t}$,   see (\ref{R4}), is a ground state of the dipole ring of length $N$.\\

\fbox{Case C}\\

\begin{figure}
\begin{center}
\includegraphics[clip=on,width=150mm,angle=0]{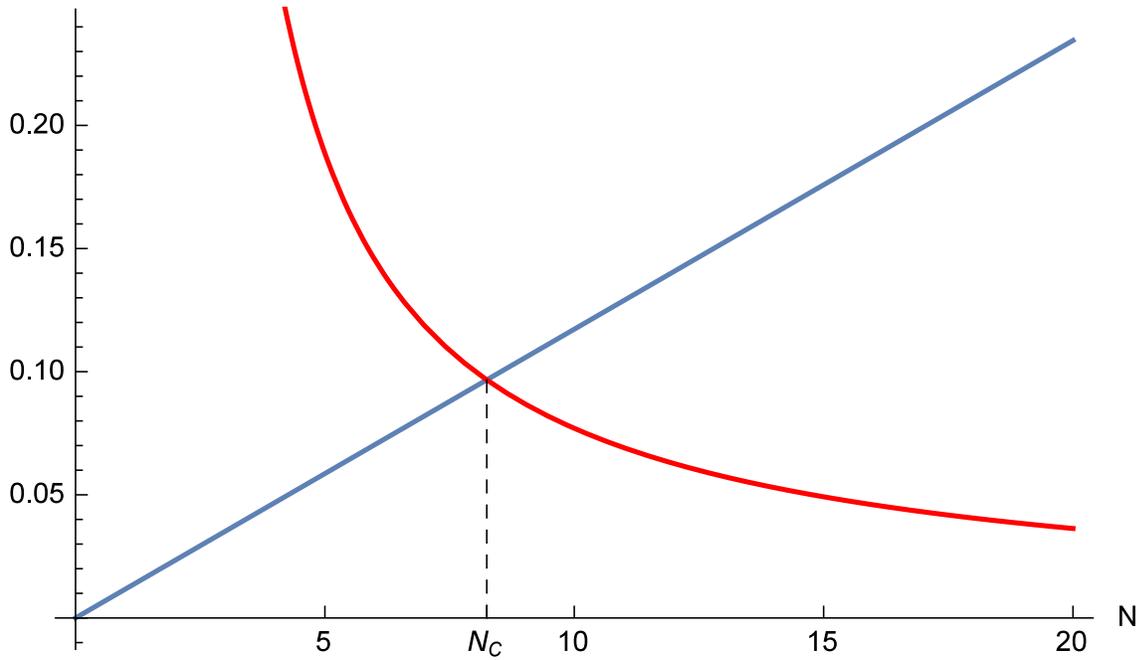}
\end{center}
\caption{Intersection of the scaled bounds  $N^{-5}B_{C1\alpha P}$ (blue graph) and $ N^{-5}B_{NC}$ (red graph) at $N_C=8.24568\ldots$ .
\label{FIGic}
}
\end{figure}

After inserting (\ref{epsilon}), (\ref{delta})
and $N_0=9$ we divide the lower bound (\ref{C1P3}) of the positive terms by $N^5$ and obtain
\begin{eqnarray}
\nonumber
  N^{-5} B_{C1\alpha P}&=&\frac{4}{\pi^6} N \sin ^2\frac{8 \pi }{41}\left(3+\cos \frac{2 \pi}{9}\right)
  \left(3+\sin ^2\frac{8 \pi }{41}    \left(\cos \frac{2 \pi }{9}-3\right)\right)\\
  &&
  \label{Bound3a}\\  \label{Bound3b}
  &&
   = 0.0117242\ldots N
    \;.
 \end{eqnarray}
This is an increasing linear function of $N$. On the other hand, the upper bound of the absolute value of the possibly negative terms
is evaluated as follows:
\begin{eqnarray}\label{Bound4a}
   N^{-5}B_{NC}&\equiv& N^{-5}\left(B_{CNs}+ B_{C2\alpha N1}+B_{C2\beta N} \right) \\
   \label{Bound4b}
   &=&\frac{1}{9 N^5}+\frac{52}{9 N^3}+\frac{9 \zeta (3)+2}{18  N}
   \;.
  \end{eqnarray}
  This is a  decreasing function of $N>0$ with the limits $0$
   for $N\longrightarrow\infty$ and $\infty$ for $N\longrightarrow 0$.
   The graphs of both functions (\ref{Bound3a}) and (\ref{Bound4b}) hence intersect at a unique point with
  $N_C=8.24568\ldots$, see Figure \ref{FIGic}. Hence for $N\ge N_\ast=9 >N_C$ we have $B_{C1\alpha P}>B_{NC}$ and
  the state $\pm {\mathbf t}$,   see (\ref{R4}), is a ground state of the dipole ring of length $N$.\\

\fbox{Case A}\\

\begin{figure}
\begin{center}
\includegraphics[clip=on,width=150mm,angle=0]{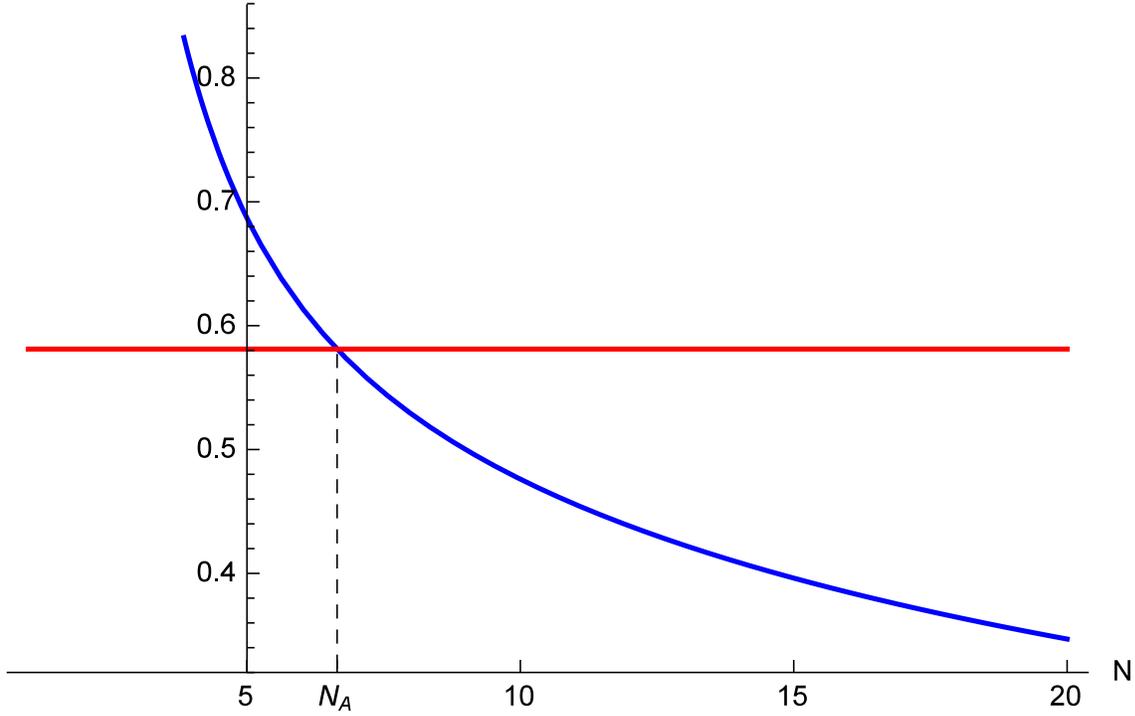}
\end{center}
\caption{Intersection of the scaled bounds  $S_0=N^{-4}B_{A1\alpha P}$ (red graph) and $ h(N)=N^{-4}B_{NA}$ (blue graph) at $N_A=6.64846\ldots$ .
\label{FIGia}
}
\end{figure}

After inserting (\ref{epsilon}), (\ref{delta})
and $N_0=9$ we divide the lower bound (\ref{LB3}) of the positive terms by $N^4$ and obtain the constant function
\begin{eqnarray}
\label{Bound5a}
 S_0\equiv N^{-4} B_{A1\alpha P}&=&\frac{825+176 \sin \left(\frac{\pi }{18}\right)-10 \cos \left(\frac{\pi
   }{9}\right)+1260 \cos \left(\frac{2 \pi }{9}\right)}{32 \pi ^4}\\
   \label{Bound5b}
  &&
   = 0.581113\ldots
    \;.
 \end{eqnarray}

 On the other hand, the scaled upper bound of the absolute value of the sum of the possibly negative terms
is evaluated as follows:
\begin{eqnarray}\label{Bound6a}
   N^{-4}B_{NA}&\equiv& N^{-4}\left(B_{A1\alpha N}+ B_{A1\beta N}+B_{A2\alpha N} +B_{A2\beta N}\right) \\
   \label{Bound6b}
   &=&SB_1+SB_2+SB_3+SB_4\;, \mbox{ where }\\
    \label{Bound6c}
   SB_1&\equiv& \frac{13}{2 N^2}\;,\\
   \label{Bound6d}
   SB_2&\equiv&\frac{13 \pi  \left(9 \log \left(\frac{N}{4}\right)+9 \gamma   +2\right)}{477 \sqrt{2} N^{5/2}}\;,\\
   \label{Bound6e}
   SB_3&\equiv&
   \frac{13 \pi  \left(9 \log \left(\frac{N}{4}\right)+9 \gamma   +2\right)}{477 \sqrt{2} N^{3/2}}\;,\\
   \label{Bound6f}
   SB_4&\equiv&
   \frac{169 \pi ^2 \left(9 \log \left(\frac{N}{4}\right)+9 \gamma
   +2\right) \left(9 \log \left(\frac{N}{4}\right)+9 \gamma   +20\right)}{227529 N}
   \;.
  \end{eqnarray}
 Consider the real function $N\mapsto h(N)\equiv N^{-4}B_{NA}$ defined for $N>0$.
We want to show that
\begin{lemma}\label{LNA}
 $h$ is a decreasing function for $N\ge 9$.
\end{lemma}

{\bf Proof}: It is obvious that all terms $SB_i,\,i=1,2,3,4$ are positive for $N>4$ and vanish for $N\longrightarrow\infty$. Hence for $i=2,3,4$
the largest zero $N_i$ of the derivative $\frac{d}{dN}SB_i$ represents a local maximum (or a saddle point) and $N\mapsto  SB_i$ is  decreasing for $N>N_i$.  For $N\mapsto  SB_1$  the  decrease is obvious. It is a straightforward task to calculate the zeroes $N_i$ and we will only give the results:
 \begin{eqnarray}\label{zero3}
   N_2 &=& 4 e^{\frac{8}{45}-\gamma }=2.68279\ldots\;, \\
   \label{zero4}
    N_3 &=&4 e^{\frac{4}{9} - \gamma }=3.50266\ldots\;, \\
   \label{zero5}
    N_4 &=&4 e^{-\frac{2}{9}+\sqrt{2}-\gamma }=7.39697\ldots\;.
 \end{eqnarray}
 This completes the proof since all $N_i<9,\,i=2,3,4$, and $h(N)$ is a sum of four
 decreasing functions for $N\ge 7.39697\ldots$ .  \hfill$\Box$\\

We calculate the value  $h(9)=0.499948\ldots<S_0$, see (\ref{Bound5b}). Hence, according to lemma \ref{LNA},
$S_0>h(N)$ for $N\ge N_\ast=9$. Numerical calculations show that there
is a zero of $h(N)-S_0$  at $N_A=6.64846\ldots$, see Figure \ref{FIGia}, and hence even $S_0>h(N)$ for $N > N_A$,
but this result will not be used in the proof.

This completes the proof that $\pm{\mathbf t}$ is a ground state for $N\ge 9$. Together with the analytical results for $N=3,\ldots,8$,
see section \ref{sec:A}, the main result of this paper is hence proven. By the present method we cannot exclude the existence of other
ground states beside $\pm{\mathbf t}$ assuming the same ground state energy $E_0$ but this seems to be extremely unlikely.

\section*{Acknowledgment}
I thank my co-autors of \cite{SSL16}, Christian Schr\"oder and Marshall Luban,
for the permission to use parts of our joint publication for the present paper and for their continuous support.
Moreover, I am indebted to Thomas Br\"ocker for valuable discussions about the subject of this article.

\end{document}